\documentclass{article}

\usepackage{arxiv}

\usepackage[utf8]{inputenc} % allow utf-8 input
\usepackage[T1]{fontenc}    % use 8-bit T1 fonts
\usepackage{hyperref}       % hyperlinks
\usepackage{url}            % simple URL typesetting
\usepackage{booktabs}       % professional-quality tables
\usepackage{amsfonts}       % blackboard math symbols
\usepackage{nicefrac}       % compact symbols for 1/2, etc.
\usepackage{microtype}      % microtypography
\usepackage{lipsum}		% Can be removed after putting your text content
\usepackage{graphicx}
\usepackage{natbib}
\usepackage{doi}

\usepackage{caption}
\usepackage{subcaption}
\usepackage{todonotes}
\usepackage{bm}
\usepackage{multirow}
\usepackage{lscape}
\usepackage{verbatim}
\usepackage[normalem]{ulem}
\useunder{\uline}{\ul}{}
\usepackage{amsmath}

\title{Generative Aging of Brain Images with Diffeomorphic Registration}

\author{ \href{https://orcid.org/0000-0003-4175-395X}{\includegraphics[scale=0.06]{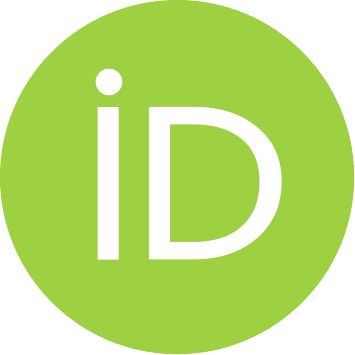}\hspace{1mm}Jingru Fu}\thanks{Corresponding author%Use footnote for providing further
		} \\
	Division of Biomedical Imaging\\
	KTH Royal Institute of Technology\\
	\texttt{jingruf@kth.se} \\
	\And
	\And 
	Antonios Tzortzakakis\\
	Division of Radiology, Karolinska Institute\\
	Medical Radiation Physics and Nuclear Medicine\\
	Karolinska University Hospital\\
	\texttt{antonios.tzortzakakis@ki.se} \\
	\And
	Jos\'e Barroso\\
	Department of Psychology\\
	University Fernando Pessoa Canarias\\
	\texttt{jbarroso@ull.es} \\
	\And
	\href{https://orcid.org/0000-0002-3115-2977}{\includegraphics[scale=0.06]{Figures/orcid.pdf}\hspace{1mm}Eric Westman}\\
	Division of Clinical Geriatrics\\
	Karolinska Institute\\
	Department of Neuroimaging\\
	King’s College London\\
	\texttt{eric.westman@ki.se} \\
	\And
	Daniel Ferreira \\
	Division of Clinical Geriatrics\\
	Karolinska Institute\\
	\texttt{daniel.ferreira.padilla@ki.se} \\
	\And
	Rodrigo Moreno \\
	Division of Biomedical Imaging\\
	KTH Royal Institute of Technology\\
	\texttt{rodmore@kth.se} \\
	\And
	for the Alzheimer’s Disease Neuroimaging Initiative\thanks{
	Data used in preparation of this article were obtained from the Alzheimer’s Disease Neuroimaging Initiative (ADNI) database (adni.loni.usc.edu). As such, the investigators within the ADNI contributed to the design and implementation of ADNI and/or provided data but did not participate in analysis or writing of this report. A complete listing of ADNI investigators can be found at: \url{http://adni.loni.usc.edu/wp-content/uploads/how_to_apply/ADNI_Acknowledgement_List.pdf}
		}
}

\date{\vspace{-5ex}}

\renewcommand{\headeright}{Fu et al.}
\renewcommand{\shorttitle}{Generative Aging of Brain Images with Diffeomorphic Registration}
\begin{document}
\maketitle

\begin{abstract}
Analyzing and predicting brain aging is essential for early prognosis and accurate diagnosis of cognitive diseases. The technique of neuroimaging, such as Magnetic Resonance Imaging (MRI), provides a noninvasive means of observing the aging process within the brain. With longitudinal image data collection, data-intensive Artificial Intelligence (AI) algorithms have been used to examine brain aging. However, existing state-of-the-art algorithms tend to be restricted to group-level predictions and suffer from unreal predictions. This paper proposes a methodology for generating longitudinal MRI scans that capture subject-specific neurodegeneration and retain anatomical plausibility in aging. The proposed methodology is developed within the framework of diffeomorphic registration and relies on three key novel technological advances to generate subject-level anatomically plausible predictions: i) a computationally efficient and individualized generative framework based on registration; ii) an aging generative module based on biological linear aging progression; iii) a quality control module to fit registration for generation task. Our methodology was evaluated on 2662 T1-weighted (T1-w) MRI scans from 796 participants from three different cohorts. First, we applied 6 commonly used criteria to demonstrate the aging simulation ability of the proposed methodology; Secondly, we evaluated the quality of the synthetic images using quantitative measurements and qualitative assessment by a neuroradiologist. Overall, the experimental results show that the proposed method can produce anatomically plausible predictions that can be used to enhance longitudinal datasets, in turn enabling data-hungry AI-driven healthcare tools.
\end{abstract}

% keywords can be removed

\section{Introduction}
%general introduction
Brain aging is usually associated with cognitive decline and an increased risk of neurological disorders~\citep{popescu2021u}. 
For example, several cognitive disorders are associated with aging, such as Alzheimer's disease (AD) and stroke \citep{ma2021primary}.
Research on brain aging can reveal the underlying spatio-temporal structure of brain changes due to aging~\citep{giorgio2010age,peters2006ageing}, which is important for the early prognosis and the accurate diagnosis of diseases such as AD~\citep{alberdi2016early,mueller2005ways}. In the last decades, the collection of longitudinal brain images has facilitated research on brain aging by enabling a noninvasive way to track brain changes and observe disease progression over time~\citep{resnick2003longitudinal}. For example, one of the most-known data-sharing initiatives, the Alzheimer’s Disease Neuroimaging Initiative (ADNI) \citep{mueller2005ways} has collected images using different modalities of subjects at different stages of Alzheimer’s Disease. Compared with other modalities, Magnetic Resonance Imaging (MRI) has superior soft-tissue contrast. This has fostered the use of MRI-based biomarkers for tracking aging or disease-related neurodegenerative changes ~\citep{njeh2008tumor,devic2012mri,schmidt2015radiotherapy,macdonald2021mri}.

%motivation of MIG  
Recent studies have demonstrated the potential of using Machine Learning (ML) techniques to study brain aging~\citep{choi2018predicting,anaturk2021prediction,ouyang2021self,popescu2021u,cole2015prediction}. However, it has rarely been straightforward to follow or analyze the age and disease progressions via those learning-based methods. On the one hand, ML-based methods, especially Deep Learning (DL)-based methods, are designed to distill knowledge from data. Therefore, the availability of ground truth data is crucial to feed the data-hungry DL models. Nevertheless, the sensitive nature of medical data makes it usually difficult to access. Moreover, brain aging research needs longitudinal data, which is even less available or unstable due to, for example, scans taken at different intervals of time or scanners. On the other hand, the high dimensionality of the brain images exponentially increases the resource demands of training DL models~\citep{wegmayrGenerativeAgingBrain2019a}. For this reason, many studies have limited their scope to generating 2-Dimensional (2D) slices extracted from the 3-Dimensional (3D) MRI scan, e.g., \citet{bowles2018modelling,kim2021longitudinal,pathan2018predictive}. Although the 3D MRI scan can be reconstructed by concatenating 2D slices, it is difficult to assess its internal consistency, with a risk of losing its anatomical plausibility. For all these reasons, it is becoming increasingly necessary to develop \textbf{Medical Image Generation (MIG)} models, which aim at generating trusted and accurate synthetic high-dimensional medical images according to the applications.
%In addition, individualized prediction and diagnosis are essential in medical imaging due to patients' distinct characteristics. 
%, efficient samples from the same individual are difficult to guarantee due to the irregular acquisition time and the amount of training data required for DL models. 
  %and may fail to be used as intended. 
%A computationally efficient and anatomically plausible MIG model that works in 3D is therefore urgently needed. 

% contribution of the proposed method
In this study, we propose a 3D MIG model based on diffeomorphic registration, aiming at synthesizing MRI scans with increasing age, in which subject-level predictions can be derived from individualized image pairs. With our model, the aging progression of the brain can be simulated rapidly with high-dimensional MRI scans. For example, it could synthesize brain atrophy progression from age 60 to age 80 as represented by MRI scans for a particular subject. The main contributions of the proposed method can be summarized as follows: i) we develop a new MIG pipeline that can synthesize subject-specific and anatomically plausible MRI series in a computationally efficient manner;
ii) we introduce an aging generative module (AGM) that does not require a training phase and can be applied to any framework that is based on diffeomorphic registration;
iii) we introduce the Quality Control Module (QCM) working in conjunction with AGM, which is used to assess the quality of the synthetic images according to the input pair; iv) we augment the existing longitudinal MRI scans with corresponding segmentations by around three times and make them public, enabling the development of data-hungry AI-driven healthcare tools, for instance, developing registration and segmentation algorithms for high-resolution predictions, which always require more data to achieve.

% brief point out the experiment results %this is more for the discussion section
%Our evaluation of the proposed method involves two aspects: 1) assessing its ability to simulate aging progression; and 2) evaluating the quality of the synthetic images.
%Our methodology was evaluated on 2622 T1-weighted MRI scans from 796 participants collected in three different datasets. First, the aging simulation ability of the proposed methodology is evaluated in two ways: i) visualization of generated aging MRI scans and the corresponding difference maps; ii) trade-off strategies by using "middle" MRI scans to give statistical evaluations; Secondly, the quality of the synthetic images is evaluated from three distinct perspectives: i) quantitative measurement; ii) qualitative measurement: we ask an expert to evaluate from clinical perspective. Overall, the experimental results show that the proposed method can produce subject-level anatomically plausible predictions.

% the structure of the remaining of the paper
% Four sections make up the remainder of the paper. The background and related work are elaborated in Sect. \ref{section:Relatedwork}. A detailed explanation of the proposed method is given in Sect. \ref{section:method}. Sect. \ref{section:experiments} shows the experimental results on three independent datasets. We discuss the performance and implications of the proposed method in Sect. \ref{section:Discussion}. Finally, Sect. \ref{section:Conclusion} outlines the main conclusions of the study.

\section{Related work}\label{section:Relatedwork}
% mainly introduce MRI techniques in the past
The aging population has increased concern for age-related neurodegenerative diseases and so, aging cohorts and studies have attracted growing interest. MRI scans can clearly illustrate the anatomical structure inside the brain and thus have been used in research on aging. To analyze aging or chronic disease progressions of the brain, AI-based MIG models have been introduced to synthesize scans at different stages of diseases or at different ages.

% talk about the GANs and the computational problem
A commonly used architecture in MIG models is Generative Adversarial Networks (GANs)~\citep{creswell2018generative}. GANs are designed to generate new data from the same distribution, which consists of two parts: a discriminator to distinguish fake and real samples and a generator to learn new plausible samples to deceive the discriminator. Several GAN-based methods have been introduced to model the aging progression, e.g., \citet{wegmayrGenerativeAgingBrain2019a,bowles2018modelling,kim2021longitudinal}. Training GANs with 3D brain images is challenging mainly due to the high dimensionality of the brain images. As a result, most of the previous studies have simplified the problem by either using only a single slice per subject \citep{wegmayrGenerativeAgingBrain2019a} or by downsampling the original images, which might result in poor resolution predictions \citep{ravi2022degenerative}. To alleviate these limitations, \citep{jung2021conditional} proposed a method to synthesize high-quality 3D medical images by introducing a 3D discriminator in a normal 2D GAN architecture. A depth-wise concatenation module was introduced to concatenate separate 2D slices into a whole 3D image. Apart from the technical challenges, the main issue of GAN-based methods is that they are unable to guarantee the anatomical plausibility of the generated images due to the lack of biologically informed constraint in the generation. This issue becomes relevant if the synthetic images are expected to be used for answering clinical questions.

%Even though this work can synthesize high-quality 3D MRI scans by the use of a set of 2D and 3D discriminators, the generated MRI scans cannot guarantee the anatomically plausible due to the lack of biologically informed constraint. 
%here I am (RM)
% another problem individualization and the regression models
Apart from the aforementioned issues when using GAN as the architecture, another critical factor for analyzing aging is individualization. Personalized healthcare and individualized medicine are important since each patient has different physical conditions that may cause the same disease. With regard to the aging-related processes, it is even more complicated since aging-related brain changes can be influenced or driven by several factors, such as Alzheimer’s disease (AD)~\citep{song2020longitudinal}, traumatic brain injury (TBI)~\citep{cole2015prediction}, even different regions of the brain might follow a different aging pattern~\citep{popescu2021u}. Image regression is therefore introduced as a means to encode personalized information in GANs.

Image regression was introduced with the aim of estimating images as a function of associated variables such as age~\citep{niethammer2011geodesic, beg2005computing}. The complexity of analyzing age or disease progressions was alleviated by modeling regression approaches at the population level~\citet{dukart2013generative,huizinga2018spatio}. For example, those group-level methods aim at simulating spatio-temporal changes during aging across all subjects. Even though these methods capture the time-varying changes of a population well, the way to leverage and extrapolate this to the target subject is still under development~\citep{campbell2017efficient}. The work of~\citet{pathan2018predictive} has addressed this extrapolation problem by incorporating the regression model based on the framework of Large Deformation Diffeomorphic Metric Mapping (LDDMM)~\citep{pathan2018predictive} with convolutional neural networks (CNN) and recurrent neural networks (RNN). Their model, however, generated a sequence of vector moments under the LDDMM framework before model training, so performance and the model were highly dependent on the LDDMM output. %Another alternative to breaking the individualization limitation is collecting more samples from the same subject, as many data-sharing initiatives do, offering a solution by establishing longitudinal medical images using anonymous patients' personal information and giving them a unique identification number. It is, however, problematic to share medical image data due to privacy concerns. Even with those efforts, the longitudinal datasets lack the ability to give personalized diagnoses due to the limited and irregular acquirements of subject-level data. 

The fundamental tool for performing image regression is diffeomorphic image registration%.
%The pattern of the paired input reminded us of a related fundamental component of medical image analysis pipelines: image registration, 
, which aims at estimating spatial correspondences between images~\citep{zitova2003image}. Traditional methods for diffeomorphic image registration are very time-consuming, which has limited the application of image regression in different contexts. Recently, many deep-learning-based diffeomorphic registration methods have emerged~\citep{fu2020deep,dalca2019unsupervised, hoffmann2020synthmorph, balakrishnan2019voxelmorph, chen2021transmorph,Li2022} with the aim of reducing the computational burden. These methods have shown a comparable registration accuracy to that of the traditional methods, especially for brain images.

% summary, and new pipeline that we found
Inspired by the aforementioned methods for simulating brain aging on longitudinal data, we found it necessary to develop a computational effective and anatomically plausible MIG model for 3D scans. Most of the GAN-based methods use one input image to generate synthetic images. Instead, we use two scans at different time points. Aging is a complex process that is influenced by many factors, such as lifestyle factors, cognitive diseases, education, etc. Using a pair of inputs from the same subject can provide a more accurate picture of the individual aging. With this feature, we can obtain images of a higher quality which can be used to augment the available longitudinal datasets.
%tend to hold the simple assumption that aging is solely dependent on the current observation, thus the predictions are independent of the subjects' age~\citep{wegmayrGenerativeAgingBrain2019a}. 
%The image regression-based method~\citep{pathan2018predictive} alleviated the single observation assumption by introducing the vector momentum sequence captured by LDDMM. However, the architecture still needs to be trained on the entire population and might cause bias on the population. 

%. We therefore propose generating aging images of the brain based on deep-learning-driven diffeomorphic image registration. Fig.~\ref{fig:fig1} shows the pipeline of the method. We address the problem of predicting aging given two images from the same subject at different time points within the framework of diffeomorphic registration. Specifically, two modules are proposed within the registration framework. The Aging Generative Module (AGM) aims at generating aging images, and the Quality Control Module (QCM) aims at updating AGM according to the quality of generated images. In comparison to group-based image regression and GAN methods, the proposed approach can better capture the growth trend of a target subject by focusing on a pair of inputs. The anatomical plausibly can be controlled in two folds. First, the paired inputs already constrain the starting and ending points within the registration framework, which determines the outputs must fall between them. Secondly, we propose the QCM for determining and updating the integration ending point based on the quality of the generated images.

\section{Method}\label{section:method}

\begin{figure}[!h]
\centering
\includegraphics[width=1\linewidth]{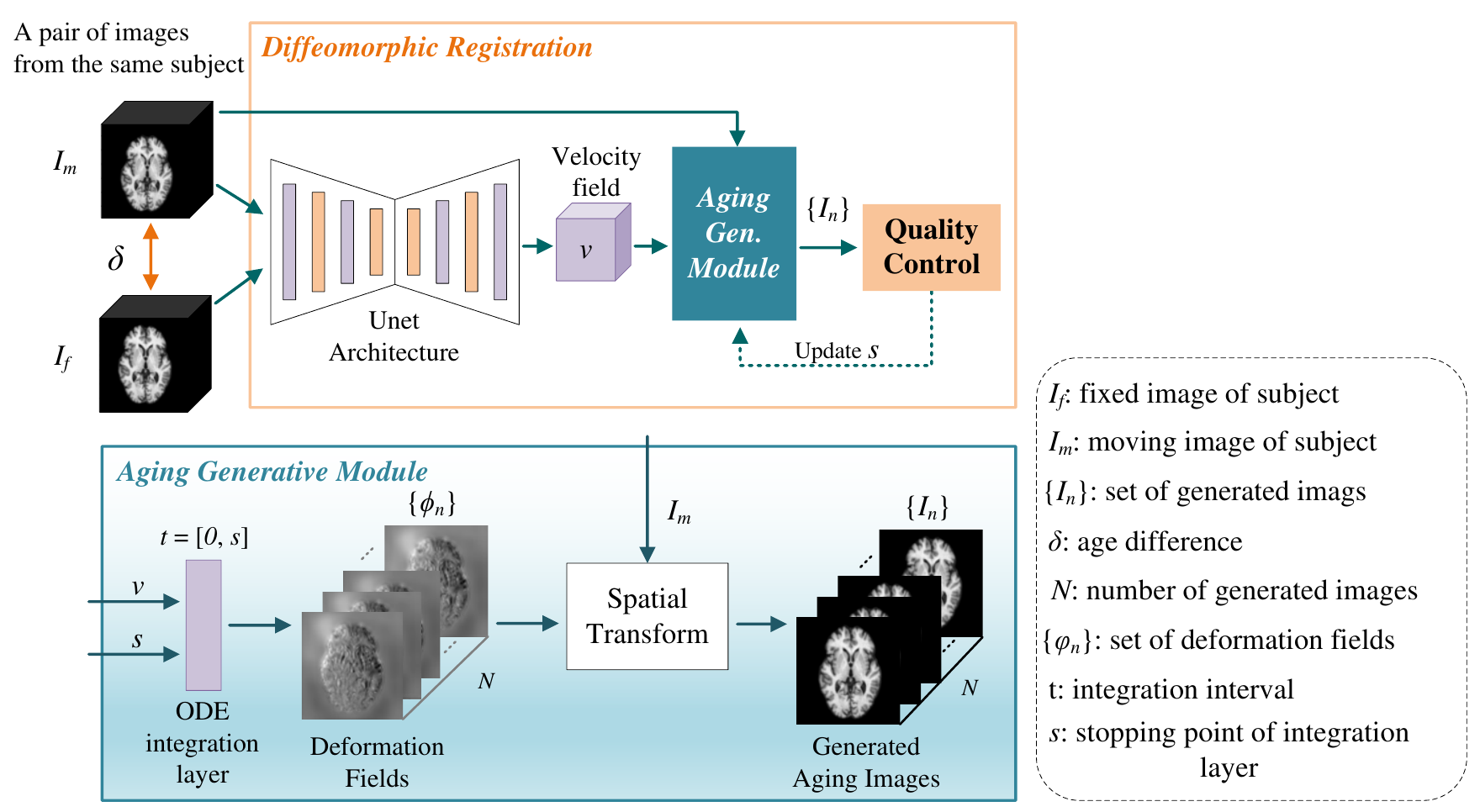}
\caption{ Architecture of the proposed Medical Image Generation (MIG) model. The input is an individualized image pair, where $I_{m}$ stands for \emph{moving} image, $I_{f}$ \emph{fixed} image. The proposed two modules take \emph{velocity field} as input. \emph{Aging Generative Module (AGM)} and \emph{Quality Control Module (QCM)} are introduced within the skeleton of diffeomorphic registration. The details of AGM are shown in the bottom blue shadow part. The \emph{deformation fields} can be derived from \emph{velocity field} given the subject-specific stopping point $s$ and corresponding number of generated MRI scans $N$. At the end, aging MRI sequences can be derived through Spatial Transform. The \emph{Quality Control Module (QCM)} can provide the subject-specific $s$ by applying quality measurements between generated and fixed MRI scans.}
\label{fig:fig1}
\end{figure}

Figure \ref{fig:fig1} shows the framework of the proposed aging generative method. As shown, we assume that the input is an image pair of the same subject acquired at different time points, the objective of aging generation is to synthesize images of aging overtime for that subject. %Under our assumption, the aging needs to be based on both baseline and follow-up images, rather than just using a single baseline image as many GAN-based aging generation methods do (e.g.,\citet{wegmayrGenerativeAgingBrain2019a}).  %see my comment above
The framework consists of three main parts as shown in Fig.\ref{fig:fig1}: 1) the skeleton of diffeomorphic registration; 2) the proposed Aging Generative Module (AGM) aiming at simulating linear aging MRI scans; and 3) the proposed Quality Control Module (QCM), which provides the subject-specific hyperparameter $s$ for AGM and that is fulfilled by imposing an accurate-preserve constraint on the synthetic MRI scans.

\subsection{Diffeomorphic Registration}

Image registration is a fundamental step for analyzing medical images either in clinics or for downstream tasks such as segmentation, regression, or classification. In its simplest form, image registration involves estimating a smooth, continuous mapping between the points in one image and those in another. Specifically, given a moving image $I_m$ and a fixed image $I_f$, the goal of image registration is to find a deformation field $\phi$ to map $I_m$ into $I_f$. 
%This is "semi" true
%There are two main categories of image registration: the small-deformation framework and the large-deformation/diffeomorphic framework. Comparatively to the small-deformation framework, 
Preserving the topology is crucial when registering biological image pairs in order to avoid the folding of tissues. Diffeomorphic registration has the advantage that it computes deformation fields that are both differentiable and invertible, which means that it can preserve the topology.
%framework has a number of elegant mathematical properties, including that the deformation field is differentiable and invertible (i.e. nonzero Jacobian determinant), thus preserving topology. 

Traditional diffeomorphic registration methods are computationally intensive. Benefiting from the vigorous development of machine learning (ML), many ML-based registration methods have been proposed to shorten the registration time in the testing phase from tens of minutes to hours for the traditional methods, to a few minutes or even seconds for the ML-based ones~\citep{fu2020deep}. Several ML-based diffeomorphic registration methods have been proposed~\citep{krebs2018unsupervised,krebs2019learning,dalca2019learning,hoffmann2020synthmorph,dalca2019unsupervised,Li2022}.%, combining the speed advantage and mathematical properties in both areas. 
We chose SynthMorph \citep{hoffmann2020synthmorph} in this study because of its good performance for brain registration.
The architecture of SynthMorph is summarized 
%those ML-based diffeomorphic registration methods can be generalized
in the upper part of Fig.\ref{fig:fig1}. One of the issues of using ML for diffeomorphic registration is that it is difficult to make sure that the learned deformation fields are diffeomorphic.
To solve this issue, SynthMorph divides the problem into two steps. First, a U Net-like neural network \citep{ronneberger2015u} is trained to learn a stationary velocity field representation, $\bm v$, following a similar approach to the Diffeomorphic Anatomical Registration using Exponentiated Lie algebra (DARTEL) method \citep{ashburnerFastDiffeomorphicImage2007}, in which a single velocity field is involved which remains constant over unit time. In a second step, this vector field is used for estimating the actual diffeomorphic deformation field by solving the ordinary differential equation (ODE):

%In principle, a differential equation could be used to describe the evolution of the deformation $\phi$ with a constant velocity field $v$ over time $t$. Then the evolution of diffeomorphism can be formulated as the following Ordinary Differential Equation (ODE):

\begin{equation}
	\frac{d\bm\phi^{(t)}}{d t} = \bm v(\bm\phi^{(t)})
\end{equation}
where $\bm\phi^{(0)}$ is initialized with an identity transform. The final deformation field $\bm\phi^{(1)}$ is obtained by integrating over unit time as follows:
\begin{equation}
 \bm\phi=\bm\phi^{(1)} = \displaystyle \int_0^1  \bm v(\bm\phi^{(t)}) {d t}. \label{eq:intlayer}
\end{equation}
From group theory, the velocity field $\bm v$ can be seen as a member of the Lie algebra, which is exponentiated in order to produce a deformation $\bm\phi^{(1)}$. The resulting deformation is a member of a Lie group: $\bm\phi^{(1)}=Exp(\bm v)$. 
% \todo{
% should I give more details of diffeomorphic registration here?
% }

% Eq. (\ref{eq:intlayer}) can be solved with different ODE solvers. The most popular one is called \emph{scaling and squaring}~\citep{ashburnerFastDiffeomorphicImage2007,arsigny2006log}. With the assumption of a constant velocity field $\bm v$, the integration involves computing new solutions after many successive small time steps. As an example, a relatively accurate solution can be obtained by using eight time steps as follows:
% \begin{equation}
% \begin{array}{c}
%  \bm\phi^{(1/8)} = \bm p + \bm v (\bm p)/8\\
%  \bm\phi^{(1/4)} = \bm\phi^{(1/8)}\circ\bm\phi^{(1/8)}\\
%  \bm\phi^{(1/2)} = \bm\phi^{(1/4)}\circ\bm\phi^{(1/4)}\\
%  \bm\phi^{(1)} = \bm\phi^{(1/2)}\circ\bm\phi^{(1/2)}
%  \end{array}
% \end{equation}
% where $\circ$ denotes the composition operation and $\bm p$ is a map of spatial locations. The main advantage of this implementation is its relative low computational cost. However, this solver is not beneficial for our purposes as discussed in the next subsection. 

Eq. (\ref{eq:intlayer}) can be solved with the Euler method, which involves calculating a new solution after a series of successive small steps $h$.
\begin{equation}
    \bm\phi^{(t+h)}=(\bm p+h \bm v)\circ\bm\phi^{(t)}
\end{equation}
where $\circ$ denotes the composition operation and $\bm p$ is a map of spatial locations. As an example, a relatively accurate solution can be obtained by using eight time-steps as follows:

\begin{align}
 \bm\phi^{(1/8)} &= \bm p + \bm v (\bm p)/8\\
 \bm\phi^{(2/8)} &= \bm\phi^{(1/8)}\circ\bm\phi^{(1/8)}\\
 \bm\phi^{(3/8)} &= \bm\phi^{(1/8)}\circ\bm\phi^{(2/8)}\\
 ...\ \ \ &\ \ \ \ \ \ \ \ \ \ \ ...\\
 \bm\phi^{(1)}\ \ \ &= \bm\phi^{(1/8)}\circ\bm\phi^{(7/8)}
\end{align}

If the number of time steps is a power of 2, then it is called \emph{scaling and squaring}~\citep{ashburnerFastDiffeomorphicImage2007,arsigny2006log}. The main advantage of this implementation is its relatively low computational cost due to the simplifying of internal points.

Motivated by this, we consider extracting the output deformation fields in the middle of the integration and applying them to the spatial transform block. The evolution between two ages associated with two input images can be simulated.

\subsection{Aging Generative Module}
Our approach for synthesizing images between the fixed and moving ones is to generate different deformation fields at different time steps between the paired input. In addition, the stopping point (i.e., $s$) of integration is set as the hyper-parameter to enable the input-specific focus. The main problem of the \textit{scaling and squaring} approach for our purpose is that the deformation field is computed at irregular time steps (i.e. power of 2) in the integration range. Moreover, it is difficult to obtain fine-grained extrapolation points beyond the original stopping point when $t=1$. Since our goal is to generate samples at more regular steps, it is beneficial to use a more standard ODE solver that allows us to generate deformation fields at any time $t$. In particular, we used the TensorFlow implementation of the ODE solver by \citep{petzold1983automatic} to obtain linear intermediate outputs. This method is just slightly more computationally expensive than the \textit{scaling and squaring} approach, so it is appropriate for our current goal.

%take traditional ODE solvers to obtain linear intermediate outputs~\citep{petzold1983automatic}. 

 %Compared with the common-used \emph{scaling and squaring}, the traditional ODE solvers proposed in \citep{petzold1983automatic} can optimize ODE at any given point, 
%thus enabling the linear aging generation. In the implementation, we demonstrate that with the proposed QCM, the proposed AGM can replace the modules after velocity filed in a pre-trained ML-based diffeomorphic registration framework directly without affecting the quality of the outputs.

The bottom part of Fig.\ref{fig:fig1} shows the aging generative module (AGM) of the proposed method. %is designed within the skeleton of diffeomorphic registration as depicted in the upper part of Fig.\ref{fig:fig1}. Note that primary task of AGM is to simulate aging progression based on a pair of images from the same subject. The details of AGM are shown in the bottom part of Fig.\ref{fig:fig1}, 
First, given the velocity field $\bm v$ estimated with the neural network, we generate $N$ deformation fields at different regular time points in the range $[0, s]$, with $s$ being a parameter. 
%encodes the difference between the input image pair, which can facilitate synthesizing more realistic images than only based on one image; 
%\textbf{ii) The moving image $I_m$:} is fed into the spatial transform block and the aging sequences are then generated; 
%and \textbf{iii) the stopping point $s$:} indicates the end point of ODE integration layer, which is updated by the Quality Control Module in Section~\ref{section:QCM}. There are two main blocks in AGM, integration block and spatial transform block. 
In a second step, we use the spatial transform block introduced by ~\citet{de2017end} to generate the images at different time points by warping the moving image $I_m$ with the estimated deformation fields.
%which enables unsupervised learning registration by introducing a spatial grid. Here we perserve the same structure and use this block to apply the deformation field into the moving image. 
Parameter $s$ is referred to as the initial \textit{stopping point}. %, which is set initially to two. With this setting, we allow extrapolation of the data up to $t=2$. 
This parameter is automatically adjusted later by the quality control module as described in the next subsection.

\subsection{Quality Control Module}\label{section:QCM}

As already mentioned, in theory $\phi^{(1)}$ should be used to map $I_m$ into $I_f$. In practice, this might not happen when using ML-based methods. While the neural networks learn the most likely vector field $\bm v$ for the input images, some inaccuracies are expected due to the fact that the testing $I_m$ and $I_f$ are, in general, not used during training. In other words, deformation fields at time points different than one can yield a better matching for registering the two images. As it will be discussed, it is important for the method to accurately estimate this stopping point since the age estimation of the synthetic images is adjusted with respect to that point.

\begin {comment}
Recall that one advantage of adopting registration to the generation task is that both the starting and ending points can be bound by true image pairs, therefore, the quality of the generated image series can be maintained. Although the starting point can be strictly controlled by moving image (i.e. the evolution starts from $\bm\phi^{(0)}$ in the diffeomorphic registration), while the ending point is optimized to stop at the fixed image alternatively (i.e. $\bm\phi^{(1)}$) which does not always hold the optimal. In other words, the proposed loss measurement used to guide the optimization of DL-based diffeomorphic registration models, such as Dice~\citep{milletari2016v}, Normalized Cross-Correlation (NCC)~\citep{kaso2018computation} and Mutual Information (MI), etc., should be data-dependent. In particular, across-modality registration tasks should use MI as a measure since it is based on statistical information rather than spatial information~\citep{guo2006multi} so it is insensitive to intensity; when the training phase has segmentation, the dice can be applied to obtain better registration performance; the NCC is more effective when mono-modality images are required for registration. We therefore relax the constraint that the integration layer must stop at 1. Furthermore, there is no controllable way to measure the quality of the generated images with the ground truth. Considering both factors, we introduce QCM to adjust the stopping point of the integration layer in the AGM to find the optimal $s$ according to some controllable criteria.
\end{comment}

In order to tackle this issue, we introduce the quality control module (QCM) whose aim is to adjust the initial stopping point $s$ of the integration layer of the AGM. Our approach is to assess which integration time point yields the most similar generated image compared to $I_f$. Based on previous studies and medical image generation literature (e.g., \citet{lei2019mri,emami2018generating,gu2019generating}), we chose six different similarity measurements between two images $I_1$ and $I_2$, namely the mean absolute error (MAE), structural similarity index (SSIM), normalized cross-correlation (NCC), peak signal-to-noise ratio (PSNR), normalized Frobenius norm (NFN), and Dice score (DSC).

\begin{figure}[!h]
\centering
\includegraphics[width=0.9\linewidth]{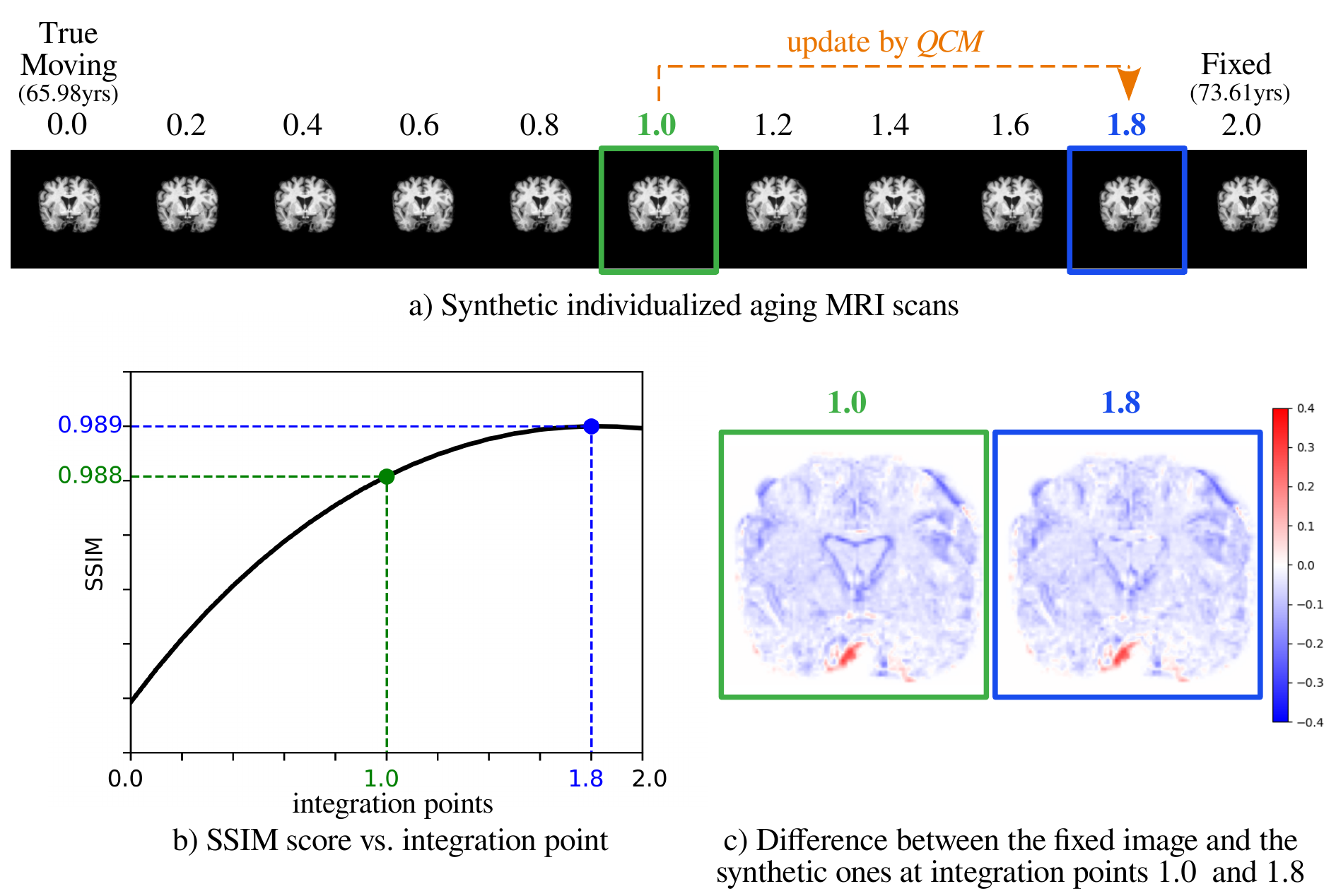}
\caption{Adjustment of the stopping point $s$. a) Series of generated images for a specific subject for different integration points. The center coronal slice from the 3D volumes is depicted in these images. b) SSIM between the fixed image and the synthetic ones is maximum at $t=1.8$. c) Local differences between the fixed image and the synthetic ones at $t=1.0$ and the optimal stopping point at $t=1.8$.}
%from one healthy subject, the stopping point is adjusted via \emph{QCM} from 1 to 1.8 in this specific example; 2) The left part shows the SSIM values w.r.t the integration points, comparing with the true fixed MRI scan showing the right side; the right part shows the difference images before and after the adjustment of \emph{QCM}. (Extracted the center coronal slice for visualization, 3D in reality)}
\label{fig:QCM}
\end{figure}

Fig.~\ref{fig:QCM} shows an example of how $s$ is adjusted for the specific case of SSIM. We have chosen hyperparameter $s$ as 2 in this case to facilitate understanding. As shown, the image at $t = 1.8$ is more similar to the fixed image than the one at $t=1.0$ according to SSIM. 
%From the quantitative perspective, the new stopping point can derive a %more similar endpoint MRI scan compared with the MRI scan obtained at old-time point 1, since the SSIM score is higher than the old one (0.988 to 0.989). 
Although the difference in SSIM is slight, the morphological differences are visible (see for example the ventricles). % differences are from the qualitative perspective shown in the right side of Fig.~\ref{fig:QCM}, the ventricle part of the brain can gain a more realistic appearance, which is vital for the aging problem. Moreover, by incorporating the quality measurement into registration, one can get rid of its data-dependence property and thus enabling our methodologies to apply for any diffeomorphic registration framework without the need for retraining. 

The six measurements are computed as follows:

%maximize a quality measureto find the optimal $s$ according to some controllable criteria. We introduced several criteria to evaluate the performance of generation in different aspects. Firstly, we review the generation targeted papers~\citep{lei2019mri,emami2018generating,gu2019generating} and summarize four commonly used criteria: Mean Absolute Error (MAE), Structural Similarity Index (SSIM), NCC, and Peak Signal-to-Noise Ratio (PSNR). These four metrics are applied between the generated images and fixed images for finding the optimal stopping point: 
\begin{equation}
\label{equ:MAE}
	MAE = \frac{1}{N}\displaystyle\sum_{i=1}^{N}|I_1(i)-I_2(i)|
\end{equation}
where $N$ is the number of voxels.
\begin{equation}
\label{equ:ssim}
	SSIM = \frac{(2\mu_1\mu_2+C_1)(2\sigma_{12}+C_2)}{(\mu_1^2+\mu_2^2+C_1)(\sigma_{1}^2+\sigma_{2}^2+C_2)},
\end{equation}
where $\mu_i$ and $\sigma_i$ stand for the mean and standard deviation of image $i$, $\sigma_{12}$ is the covariance, and the parameters $C_1=(k_1Q)^2$ and $C_2=(k_2Q)^2$ are used to stabilize divisions with weak denominators, with $Q$ being the dynamic range of the MRI scans. We used $k_1=0.01$ and $k_2=0.02$ in the experiments.
\begin{equation}
\label{equ:ncc}
	NCC = \frac{\left|\displaystyle\sum_{j=1}^{N} (\hat{I_1}(j) \hat{I}_2(j))\right|}{\left[\displaystyle\sum_{j=1}^{N} \hat{I_1}^2(j) \displaystyle\sum_{j=1}^{N} \hat{I_2}^2(j)\right]^{1/2}},
\end{equation}
with $\hat{I}_i(j)=I_i(j)-\mu_i$.
\begin{equation}
\label{equ:psne}
	PSNR = 10\log_{10}\left(\frac{Q^2}{MSE}\right),
\end{equation}
with $Q$ being the dynamic range of the MRI scans and MSE the mean squared error between the two images.

%$N$ denotes the number of voxels in Eq.~\ref{equ:MAE}. MAE shoes the absolute difference of two images, and a lower MAE value indicates better results. In Eq.~\ref{equ:ssim}, $x$ and $y$ indicate two different images respectively, while $\mu_x$/$\mu_y$ stand for the mean value of corresponding image $x$/$y$. $\delta_x$/$\delta_y$ is the variance of corresponding images. The parameters $C_1=(k_1Q)^2$ and $C_2=(k_2Q)^2$ are used to stabilize divisions with weak denominators, where $k_1=0.01$ and $k_2=0.02$. In Eq.~\ref{equ:ncc}, A and B stand for the two different images. In Eq.~\ref{equ:psne}, $Q$ is the maximum intensity value of two images, and MSE is the mean squared error. For both SSIM and PSNR, higher values indicate better prediction. 
Further, we use the normalized Frobenius Norm (NFN) (also known as the sum of squared differences)~\citep{van1996matrix} between the two images:
\begin{equation}
\label{equ:fro}
	NFN =\sqrt{ \frac{1}{N}\displaystyle\sum_{i=1}^{N}|I_1(i)-I_2(i)|^2}
\end{equation}

%It is also important to incorporate segmentation in the registration process in order to improve performance. 
Whenever segmentation masks are available, the Dice score is applied to the segmentation. In order to get segmentation masks for the generated images, the same estimated deformation fields are used to warp the segmentation masks of the moving image.
%The same deformation field is extracted and applied to the segmentation mask, 
In this case, nearest-neighbor interpolation is used instead of linear used in the spatial transform. The formula is as follows:
\begin{equation}
\label{equ:dice}
	DSC =  \frac{2\times (A\cap B)}{A+B} 
\end{equation}
with $A$ and $B$ being the two segmentation masks. The values range from 0 to 1, 1 representing a perfectly overlapping segmentation.

Lastly, we combine the six similarity measurements by computing the mean updated $s$ of the individual methods.

% Lastly, we apply MEAN and GMEAN~\citep{classics1956statistical} to combine the above criteria. Each operator is applied to all criteria other than Dice since segmentations are usually difficult to obtain.

% \todo{add an algorithm}
Once the stopping point $s$ is adjusted, we can re-generate the synthetic images with this more accurate input-specific hyperparameter.
% in order to have one image every half a year between the real ages of the moving and fixed images.

\subsection{Age estimation }\label{subsec:age_estimation}
When conducting research on aging, age is a valuable piece of information. Once the images are synthesized, the next step is to estimate the age of every synthetic MRI scan. This step is important in order to match the synthesized images with real ones by age.

It is vital to know how anatomy changes with age when it comes to the age estimation of synthetic images. \citet{walhovdEffectsAgeVolumes2005} showed that the contraction of brain structures is linear with age. Moreover, \citet{dukart2013generative} found linear decreasing age-related changes in one voxel considering GM volume at the age of 50 years as a baseline. Based on these findings, we assume that a linear increase in the integration time will lead to a linear change in the brain structures. Thus, the age of the synthetic image at time $t$, $I_t$, is computed as:
\begin{equation}
    Age(I_t)= Age(I_m) + \frac{t}{s} [Age(I_f)-Age(I_m)] .
\end{equation}

Notice that this age estimation depends on the stopping point $s$, which can be different depending on the applied measurement from the previous section. Since there are subjects with more than two acquired images in the datasets, it is possible to use the intermediate acquisitions to assess the error in the age estimation. With this, it is possible to determine which measurement is more appropriate for simulating aging with the proposed methodology.

\section{Experimental Results}\label{section:experiments}

\subsection{Datasets}
\begin{table}[t]
\centering
\caption{Summary of the datasets.}
\begin{tabular}{c|cc|ccc|cc}
\hline
&\multicolumn{2}{c|}{Complete dataset}&\multicolumn{3}{c|}{Selected images}&\multicolumn{2}{c}{Synthetic images}\\
dataset& \# Images & \# Healthy & \# Sessions \textgreater 1 & \# Subjects & Age range &\# Images & Size increase\\ \hline
ADNI      & 5,097      & 1,489              & 1,393    & 347 & 59$\sim$95 & 2,500 & 179\%                                \\
OASIS-3   & 2,044      & 1,310              & 1,066    & 353 & 42$\sim$95 & 3,948 & 370\%                              \\
GENIC     & 539       & 406               & 203     & 96  & 34$\sim$79 &1,100 & 542\%
\\ \hline
Total&7,680&3,205&2,662&796&34$\sim$95 & 7,548 & 284\%
\\ \hline
\end{tabular}
\label{tab:tab1}
\end{table}

We evaluated the generative performance of the proposed methodology on three datasets: two public available datasets, the Alzheimer's Disease Neuroimaging Initiative (ADNI)~\citep{jack2008alzheimer} and the Open Access Series of Imaging Studies (OASIS-3) dataset~\citep{lamontagne2019oasis}; and our own dataset, named GENIC. They all are 3D brain-MRI datasets. We focus only on the T1-w MRI scans in this study. 
% The purpose of ADNI~\footnote{http://adni.loni.usc.edu/about/} is to develop biomarkers for the early detection and tracking of AD. 
The ADNI ~\footnote{http://adni.loni.usc.edu/about/} was launched in 2003 as a public-private partnership, led by Principal Investigator Michael W. Weiner, MD. The primary goal of ADNI has been to test whether serial magnetic resonance imaging (MRI), positron emission tomography (PET), other biological markers, and clinical and neuropsychological assessment can be combined to measure the progression of mild cognitive impairment (MCI) and early Alzheimer’s disease (AD).
OASIS-3~\footnote{https://www.oasis-brains.org} is a retrospective compilation of data from more than 1,000 participants, including 609 cognitively normal adults and 489 individuals at various stages of cognitive decline. It contains more than 2,000 MR sessions and includes T1-w scans, among other sequences. GENIC is a population-based prospective longitudinal study from the Canary Islands in Spain, which it was started in 2004 and is currently on-going~\citep{machado_proposal_2018, nemy_cholinergic_2020}. It includes T1-w scans, among other sequences. 

%\subsection{Setup}

\textbf{Data setup.} First, FreeSurfer~\citep{fischl2012freesurfer} was applied to all datasets. Image processing and data management of ADNI and GENIC were done in theHive database system~\citep{10.3389/fninf.2013.00049}, while OASIS FreeSurfer data was obtained from \url{https://www.oasis-brains.org/#access}. FreeSurfer performs skull-stripping and bias field correction. After that, we affine registered the images into FreeSurfer's Talairach space using the \emph{talairach.xfm} atlas transform generated by recon-all. 
Affine registration is necessary since we adopt the stationary velocity model, in which the evolution of the diffeomorphism is not invariant with respect to the affine transformations~\citep{ashburnerFastDiffeomorphicImage2007}.
In order to harmonize medical data for the DL-based architecture, it is important to resample the intensity of images to a common shape and scale between 0 and 1. We also cropped the images to [160,160,192] in the experiments.

Many neurodegenerative diseases can affect brain aging~\citep{popescu2021u}. For example, it has been previously shown that the brains of patients with Alzheimer's disease (AD) tend to look older than the brains they would have expected when healthy~\citep{franke2012brain, popescu2020nonlinear}. Based on this, it is reasonable to separate healthy patients from diseased patients, especially for age estimation. 
Therefore, we only used images of cognitively healthy subjects, resulting in 1,489 images in ADNI, 1,310 images in OASIS-3, and 406 in GENIC. Further, the proposed methodology requires images acquired in at least two time points as input, so subjects with sessions fewer than two were excluded. Then 1,393 images were left in ADNI, 1,066 images in OASIS-3, and 203 images in the GENIC dataset. Details about the data included in the experiment appear in Table~\ref{tab:tab1}.

\subsection{Image generation}

\begin{figure}[t]
\centering
\includegraphics[width=1\linewidth]{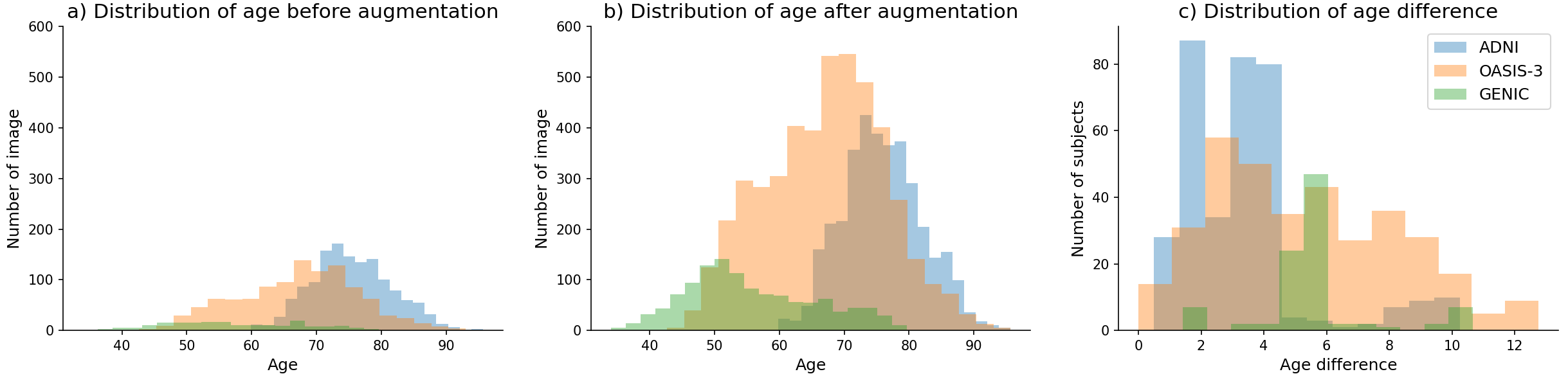}
\caption{Age distributions in the three datasets. a) and b) show the histogram of age before and after augmentation. c) shows the distribution of age difference between the first and last acquired MRI scan per subject.}
\label{fig:datasets}
\end{figure}

As mentioned, SynthMorph~\citep{hoffmann2020synthmorph} was used as the backbone of the diffeomorphic registration due to its state-of-the-art performance for DL-based diffeomorphic registration. We used pre-trained weights which were trained with a set of brain-anatomy label maps \emph{(sm-brains)}
%The registration framework applied the pre-trained weight which was trained on % random combinations of 100 T1-w and T2w images
\footnote{\url{https://surfer.nmr.mgh.harvard.edu/ftp/data/voxelmorph/synthmorph/brains-dice-vel-0.5-res-16-256f.h5}}. The number of images generated for each subject, $N_i$, is determined by the age difference (i.e., $\delta_i$ shown in Figure.~\ref{fig:fig1}) between the youngest and oldest sessions in the dataset, namely $N_i=2 \times \delta_i$ for each subject $i$. We chose the two MRI scans with the greatest age gap from a subject for two reasons: i) it will result in the longest aging simulation and relatively large augmented data pool that can be used for developing data-hungry AI-enable tools, such as registration and segmentation of 3D MRI scans; ii) the remaining intermediate MRI scans can be used for the evaluation part. It was chosen here to use double age difference because we consider the acquisition of longitudinal data to be suitable every 6 months. The initial stopping point $s$ is set as three in the experiments. The summary of the datasets and the generated synthetic images can be found in Table~\ref{tab:tab1}. It is worth mentioning that the original datasets can be augmented with high-quality MRI scans by 284$\%$. 

Fig. \ref{fig:datasets} shows the age distributions of the three datasets before and after the generation of synthetic data. As shown, GENIC contains younger subjects, ADNI older, and OASIS-3 subjects in the middle. The figure also shows that OASIS-3 covers a larger range of age differences between the first and last MRI acquisition compared to the ADNI, with GENIC in between. These differences determine that the number of synthetic images per subject in ADNI is on average smaller than in the other two datasets.

%We show the distributions of the age difference, age before and after augmentation in Figure~\ref{fig:datasets}. As we can see from the comparison between b) and c)  in Figure~\ref{fig:datasets}, by applying 6 months as the interval over 3 datasets, we augment the original datasets with around 2,500  available MRI scans to over 10,000, which is over 4 times.

%\subsection{Age estimation}

\subsection{QCM validation}

\begin{figure}[t]
\centering
\includegraphics[width=1\linewidth]{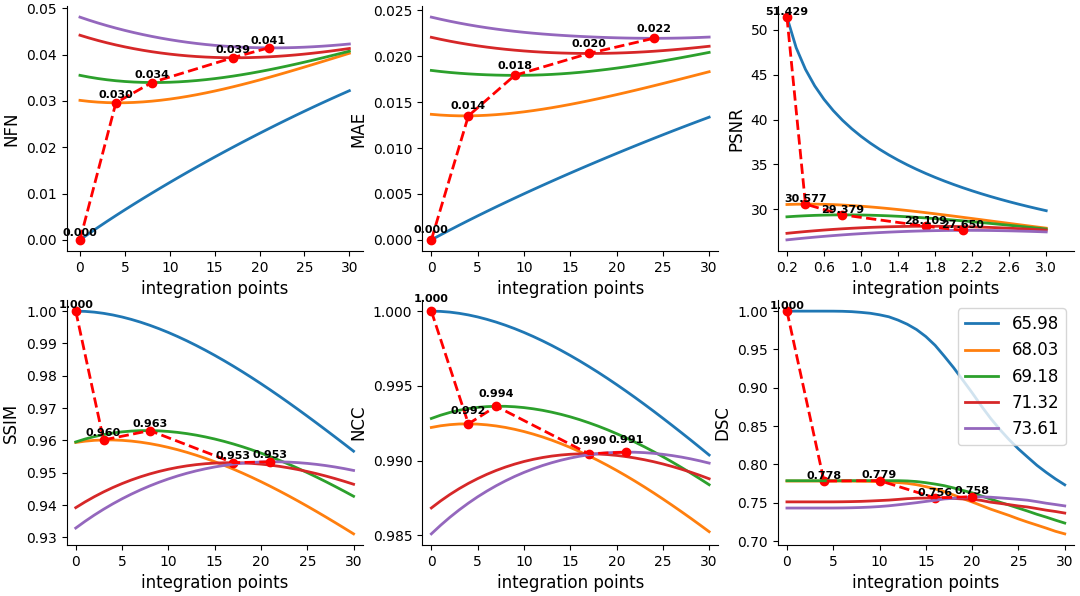}
\caption{Trends of the six quality criteria for five images of the same subject. The legends indicate the \emph{true ages} of the corresponding real MRI scans. For each curve, we calculate the "extreme" values and positions and connect them to a dashed red line. For NFN and MAE, the generated MRI scans that are most similar to the real ones are at the point of minimum value; for other measurements, it is at the maximum point.}
\label{fig:curves}
\end{figure}

Introducing QCM is one of the contributions of this work since it can take the quality of synthetic MRI scans into consideration by adjusting stopping points in the AGM at the inference phase, thus mitigating the effects of domain shift between training and test cohorts. Six similarity measurements are introduced in the QCM. The validity of QCM is evaluated from two perspectives: i) Selected measurements reflect aging-related changes; ii) The 'optimal' stopping point is mostly beyond the fixed one (i.e., $t=1$).

As discussed, the stopping point can be different depending on the quality measures used for comparing the acquired images with the synthetic ones. To show this, we randomly chose a subject from OASIS-3 that was scanned five times and used them to assess their corresponding closest synthetic images according to the different criteria. 
Every curve in Fig.~\ref{fig:curves} shows the evolution of the different measurements for the five images with the integration points.
Notice in Fig.~\ref{fig:curves} that the integration point of the closest synthetic image is always growing with the age of the acquired image for all measurements. Similar behavior was observed in subjects with more than two acquired images. This means that the chosen quality measurements are consistent and can capture aging-related changes.

\begin{figure}[t]
\centering
\includegraphics[width=1\linewidth]{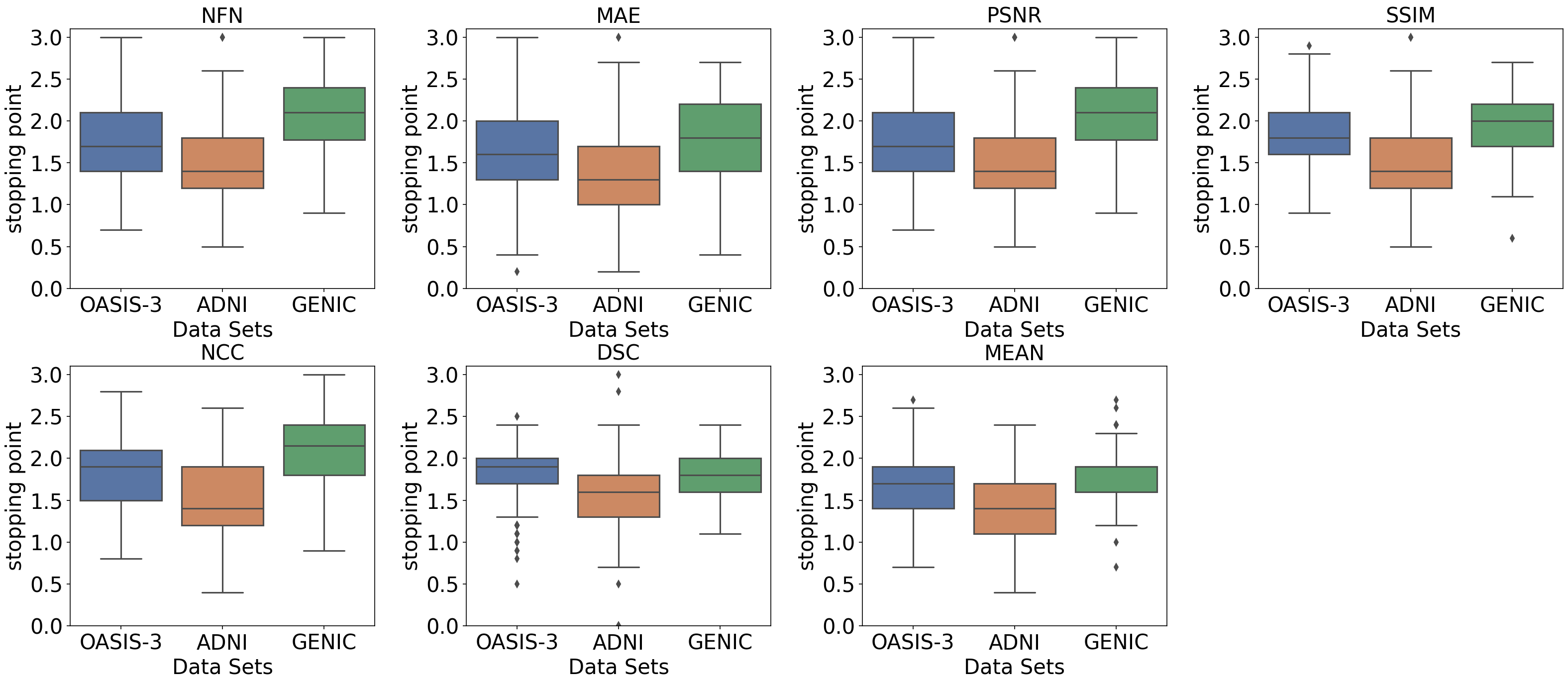}
\caption{Distribution of the stopping point per dataset for the tested quality measurements.}
\label{fig:plot_s}
\end{figure}

Figure \ref{fig:plot_s} shows box plots of the adjusted stopping points $s$ per dataset for the different quality measurements. As shown, the stopping point is higher than 1.0 in the vast majority of the cases. ADNI tends to have lower values of $s$ compared to the other datasets. This can be related to the fact that the age difference between the youngest and oldest image in ADNI was smaller than in the other datasets. Notice that the value of $s$ is relatively similar for all quality measurements.

%- \emph{Introduce RMSE as the quantitative measurement.} 
\subsection{Validation of age estimation}
At this point, it is not clear which quality measurement is the most appropriate for age estimation. To answer this question, we used the true age of the intermediate images, which were not used in the image generation, as ground truth to test the error in the age estimation.
%After assigning linear ages for middle scans, we can then compare the assigned linear ages with the true ages, then quantitative results can be obtained. 
We used the root mean square error (RMSE) between the true ages and the estimated ages as follows:
\begin{equation}
    RMSE = \sqrt{\frac{1}{N}\sum_{i=1}^{N}(Age_{true}(I_i)-Age_{estimated}(\hat{I_i}))^2}
\end{equation}
where $N$ is the number of images in the ground truth, $I_i$ is the $i$-th image in the ground truth, and $\hat{I_i}$ is its closest synthetic image according to the tested quality measurement. 

\begin{figure}[t]
\centering
\includegraphics[width=1\linewidth]{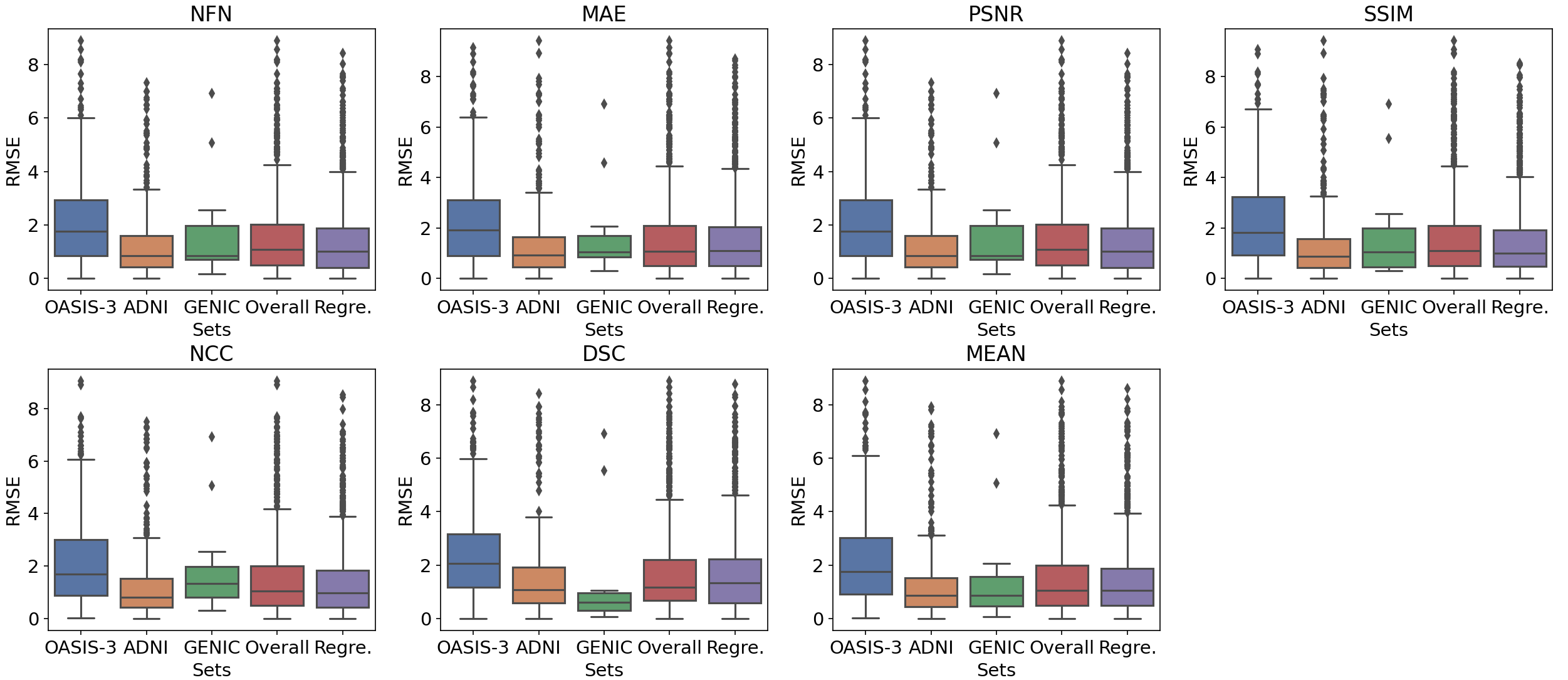}
\caption{Box plots of the RMSE of the age estimation for the tested quality measurements.}%, and it ranges from 0 to 1, the higher the better.}
\label{fig:RMSE_box}
\end{figure}

Table \ref{tab:quantitative_measures} shows the RMSE for the tested quality measurements. The corresponding box plots are shown in Fig.~\ref{fig:RMSE_box}.
%We show the results on three datasets and the mixed whole data as in Table~\ref{tab:quantitative_measures}, from which 
We observe that the RMSE is around two years when the three datasets are combined. Again, the error is lower for ADNI. As shown, NFN, PSNR, and NCC are the best options for age estimation. 
%The value of RMSE shows the deviation of linear age and true age to some extent, and it also illustrates the reliability of the assumption of linear changing of brain aging. For example, 2 means the deviation is around 2 years. But we argue that the value itself cannot directly be treated as a golden standard to say the linear assumption is bad since there is an average 2 years difference, because of two reasons, 
Notice that the RMSE estimations might be affected by quantization errors since we generate images to simulate increments of 0.5 years of age.
% and the other is that the values of criteria in a large range near the extreme value are relatively stable in Fig.~\ref{fig:curves}. 

\begin{table}[]
\centering
\caption{RMSE of the age estimation for the tested quality measurements. Columns 1 to 4 show the error of the linear model, while the last one shows the RMSE of the fitted regression lines of Fig. \ref{fig:linear_fit}.}
\label{tab:quantitative_measures}
\begin{tabular}{c|cccc|c}
\hline
%\multirow{2}{*}{\textbf{Measurement}} & \multicolumn{5}{c}{\textbf{Root Mean Squre Error (Linear age vs. True age)}}           \\ \cline{2-6} 
Measurement
                                  & OASIS-3       & ADNI    & \multicolumn{1}{c|}{GENIC}         & Three datasets & Regressions of Fig. \ref{fig:linear_fit} 
\\ \hline
NFN                          &  2.78  &  1.67  & \multicolumn{1}{c|}{2.80}          & \textbf{2.13} & \textbf{2.02}   \\
MAE                                & 2.94          & 1.84          & \multicolumn{1}{c|}{ 2.69 } & 2.28 & 2.18\\
PSNR                               &  2.78  &  1.67  & \multicolumn{1}{c|}{2.80}          & \textbf{2.13} & \textbf{2.02}\\
SSIM                               & 2.93          & 1.80          & \multicolumn{1}{c|}{2.88}          & 2.26 & 2.14\\
NCC                                & 2.80 & 1.67 & \multicolumn{1}{c|}{2.83}             &\textbf{2.13} & \textbf{2.02}\\
DSC                                & 3.00          & 1.92         & \multicolumn{1}{c|}{2.73}           & 2.35 & 2.27\\ 
MEAN                               & 2.83     & 1.72 & \multicolumn{1}{c|}{ 2.72 }         & {\ul  2.17} & {\ul 2.10}\\ \hline
\end{tabular}
\end{table}

\begin{figure}[t]
\centering
\includegraphics[width=1\linewidth]{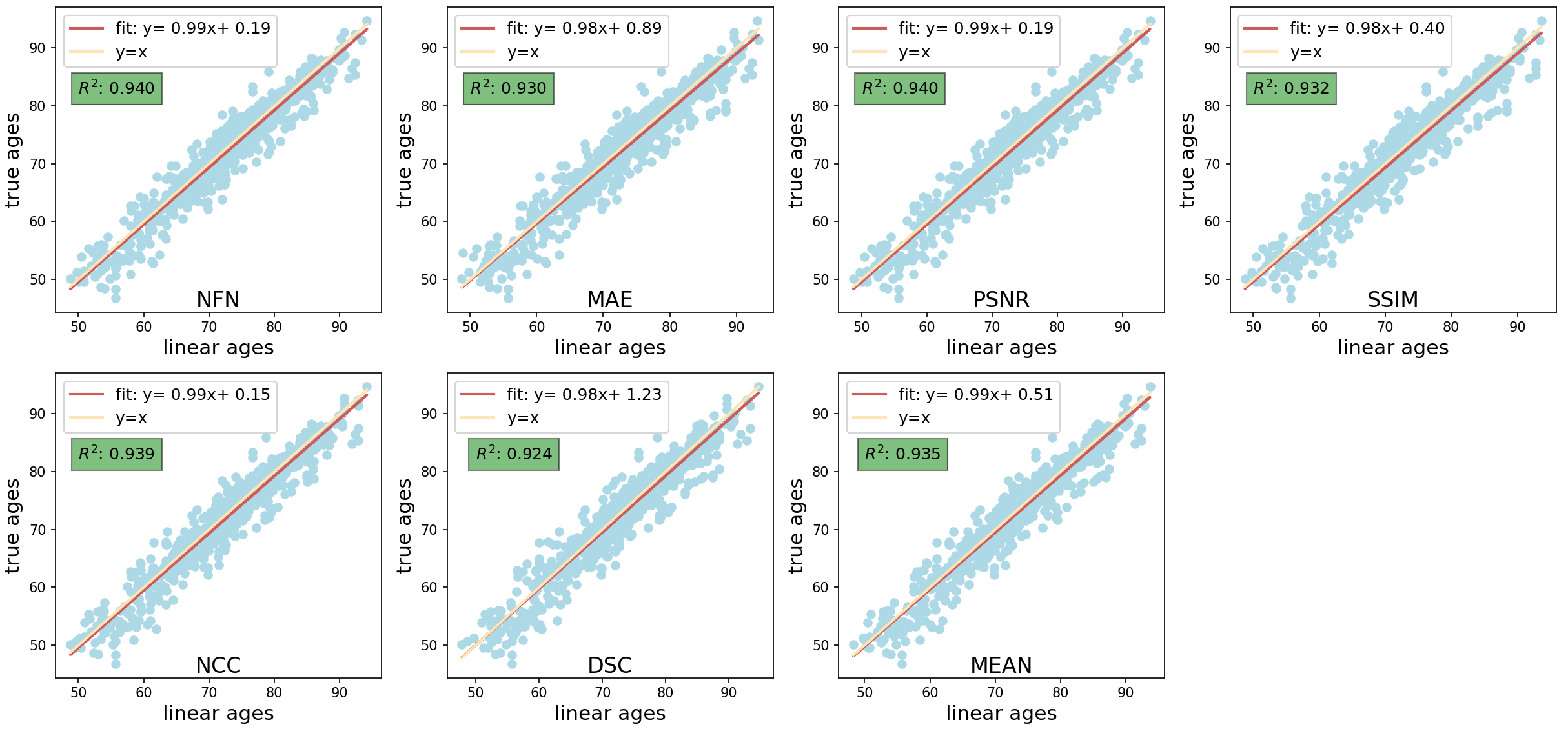}
\caption{Correlation plots between the estimated linear ages and the true ages of the ground truth images. The coefficient of determination $R^2$ is used to determine the performance of the fitting.}%, and it ranges from 0 to 1, the higher the better.}
\label{fig:linear_fit}
\end{figure}

%-\emph{Use linear regression on whole data.} 
As mentioned, previous studies have found that the brain changes linearly with age~\citep{walhovdEffectsAgeVolumes2005,dukart2013generative}. In order to assess the validity of that hypothesis in our datasets, we performed linear regressions between the real age of the images used as ground truth and the estimated as described in Subsect. \ref{subsec:age_estimation}. Fig. \ref{fig:linear_fit} shows these plots for the tested quality criteria. 
%is Assessment of the hypolinear aging To evaluate the reliability of linear assumption, we introduce the linear regression upon the obtained linear ages and the true ages for middle scans. The results as shown in Fig.~\ref{fig:linear_fit}, we use a linear function to fit them, the fitted function for each group is shown in the legend of the plot. We hope that the linear age can show a linear relationship with the real age. In another word, we hope that each linear age $x$ can directly correspond to the same value of the real age, that is, the curve we want to fit is $y=x$. To provide this guideline we further introduced the line $y=x$ as shown in the plots. We can observe that the fitted function is almost the one we want. We also introduce the coefficient of determination $R^2$, which is normally used to show either the prediction of future outcomes or the testing of hypotheses. In our case, the value of $R^2$ indicates how good the linear fitting is. 
According to the coefficient of determination $R^2$, the linear regression is valid since all measurements are higher than 0.9, which usually indicates a strong correlation between variables thus demonstrating the goodness of this fit. Ideally, the fitting lines should be $y=x$. As shown, the slopes are close to one in all cases, which validates the hypothesis that the linear changes within the brain will cause linear age increments. However, the intersect varies between 0.15 (i.e., 1.8 months) and 1.23 (i.e., one year and 2.8 months). In order to assess the effect of the intersect in the estimation of age, we added an extra column in  Table~\ref{tab:quantitative_measures} where the regression was used instead of the linear estimation of age. The improvement is in the range of 0.07 to 0.12, which is equivalent to 0.84 to 1.44 months.

%Those findings give more confidence to our linear ages assigned to generated scans. Moreover, when we get the fitted function, we can use it to "finetune" the linear ages, which gives us a better age estimation with lower RMSE in Table~\ref{tab:quantitative_measures}.

%\subsection{Quality assessment}
%We assessed the quality of the generated MRI scans from three different perspectives. i) visual inspection, ii) quantitative assessment: we calculate the statistical numbers from separate datasets to show the general quality of the generated data; iii) Clinical assessment: we ask an expert to assess the quality of the generated images and perform the discrimination task on real and synthetic MRI scans.
%: as shown in Fig.~\ref{fig:AGING}, the generated MRI scans are similar to the actual baseline from visualization since we applied diffeomorphic registration, which preserves the anatomy.
\subsection{Quantitative quality assessment}

\begin{figure}[t]
\centering
\includegraphics[width=1\linewidth]{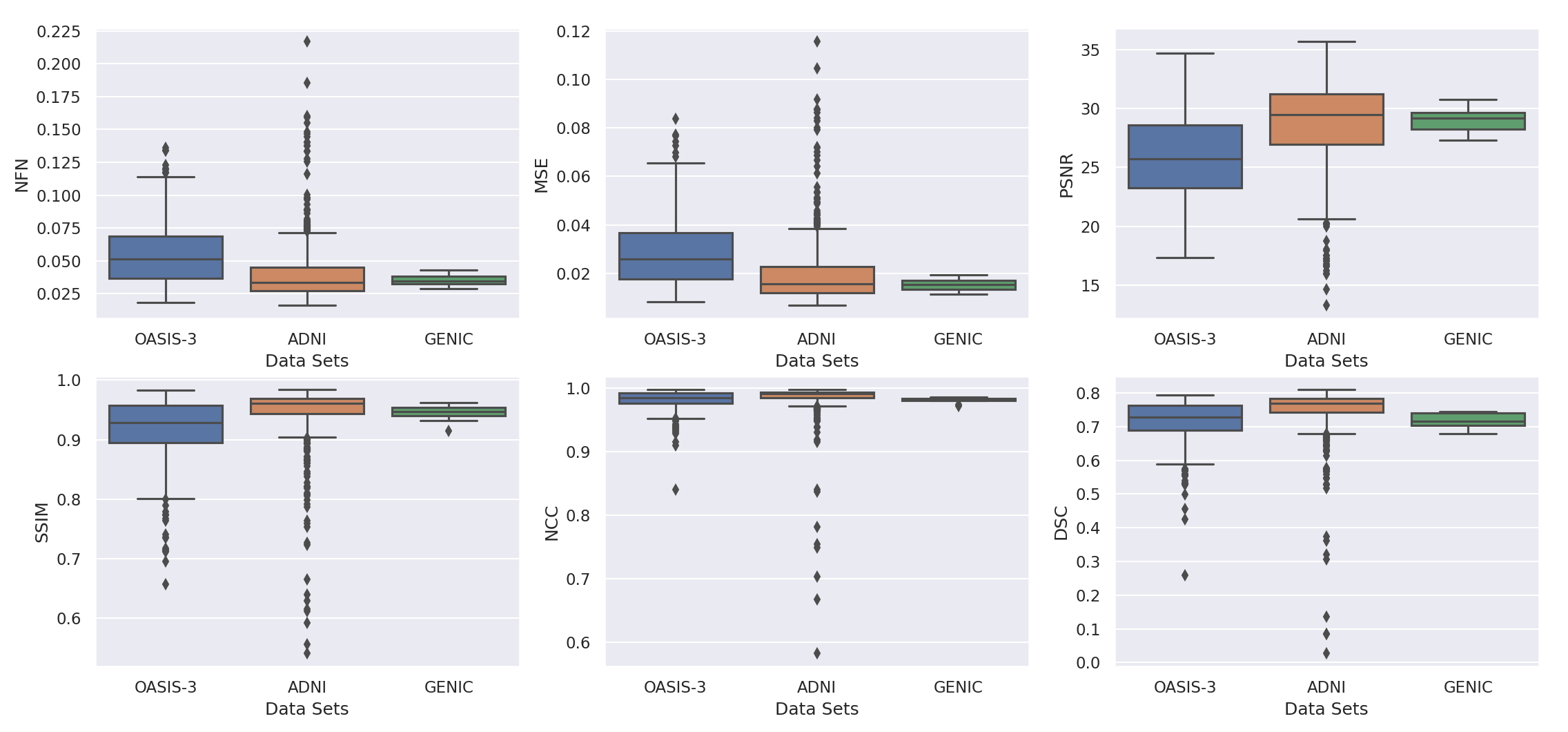}
\caption{Box plots of the quality meaurements of the most similar generated images compared to the ground truth images for different
datasets and criteria.}
\label{fig:critria_boxplot}
\end{figure}

\begin{table}[h]
\centering
\caption{Quality measurements of the most similar generated images compared to the ground truth images for different datasets and criteria. }
\label{tab:quantitative_measures1}
\begin{tabular}{c|cccc}
\hline
%\multirow{2}{*}{\textbf{Criteria}} &\multicolumn{4}{c}{\textbf{Quantitative MRI scans Quality Measures}}             \\ \cline{2-5} 
Criteria
                                  &OASIS-3        & ADNI            & \multicolumn{1}{c|}{GENIC}          & Average \\ \hline
NFN                          & 0.056         & 0.040          & \multicolumn{1}{c|}{\textbf{0.035}} & 0.044   \\
MAE                                & 0.029          & 0.020 & \multicolumn{1}{c|}{\textbf{0.016}}         & 0.022  \\
PSNR                               & 25.81          & 28.73           & \multicolumn{1}{c|}{\textbf{29.07}} & 27.87   \\
SSIM                               & 0.917          & \textbf{0.946}  & \multicolumn{1}{c|}{0.945}          & 0.936   \\
NCC                                & 0.981 & \textbf{0.985}          & \multicolumn{1}{c|}{0.981}          & 0.982   \\
DSC                                &0.715 &  \textbf{0.749}          & \multicolumn{1}{c|}{0.718}          & 0.727 \\ \hline
\end{tabular}
\end{table}

Table.~\ref{tab:quantitative_measures1} and Fig.~\ref{fig:critria_boxplot}, show the different quality measurements computed on the generated images that are most similar to the ground truth images. %on the MEAN group for each dataset separately. Note that all of them are calculated between the "middle" MRI scans and the most similar generated MRI scan. We introduced Normalized FN (NFN) in this Table to make it more comparable and interpretable. NFN is defined as the FN divided by $\sqrt{N}$, the number of the voxel in an MRI scan. So that the maximum of NFN should be normalized as 1. We also show the distribution of 6 criteria on middle scans as shown in Fig.~\ref{fig:critria_boxplot}. 
The obtained values cannot be directly compared with other generative models from the literature because of different experimental settings. Still, these values are competitive or superior to those from previous studies (e.g., %However, these values can be used for getting an But we can build a good performance intuition by comparing the number to some SOTA generation papers 
\citet{lei2019mri,emami2018generating,gu2019generating}). For example, in these studies, PSNR was around 28 (we got 27.87), SSIM was around 0.85 (we got 0.936), and NCC was around 0.93 (we got 0.982). We observed that NFN and MAE are very small as well.  

Regarding the DSC, \citet{hoffmann2020synthmorph} reported values around 0.75, compared to the mean of 0.727 of our experiments. DSC is commonly used for assessing the performance of image segmentation methods. In such applications, DCS values of 0.70-0.75 are not considered accurate. However, it is important to consider that many brain structures are small, which usually have a direct impact on the DSC. Considering that, \citet{hoffmann2020synthmorph} used the 26 large brain structures for computing the DSC.
%Likewise, in diffeomorphic registration~\citep{hoffmann2020synthmorph} the DSC can reach around 0.75 (we get 0.727). However, papers of diffeomorphic registration tend to merge smaller brain structures into larger ones in order to obtain a higher DSC (e.g. 26 structures are used in \citep{hoffmann2020synthmorph}). 
We decided to report DSC considering 113 structures in total, on Table \ref{tab:quantitative_measures1}, to get a more comprehensive overview of the performance estimated with DSC. 

\subsection{Qualitative quality assessment}

\begin{figure}
\centering
\includegraphics[width=1\linewidth]{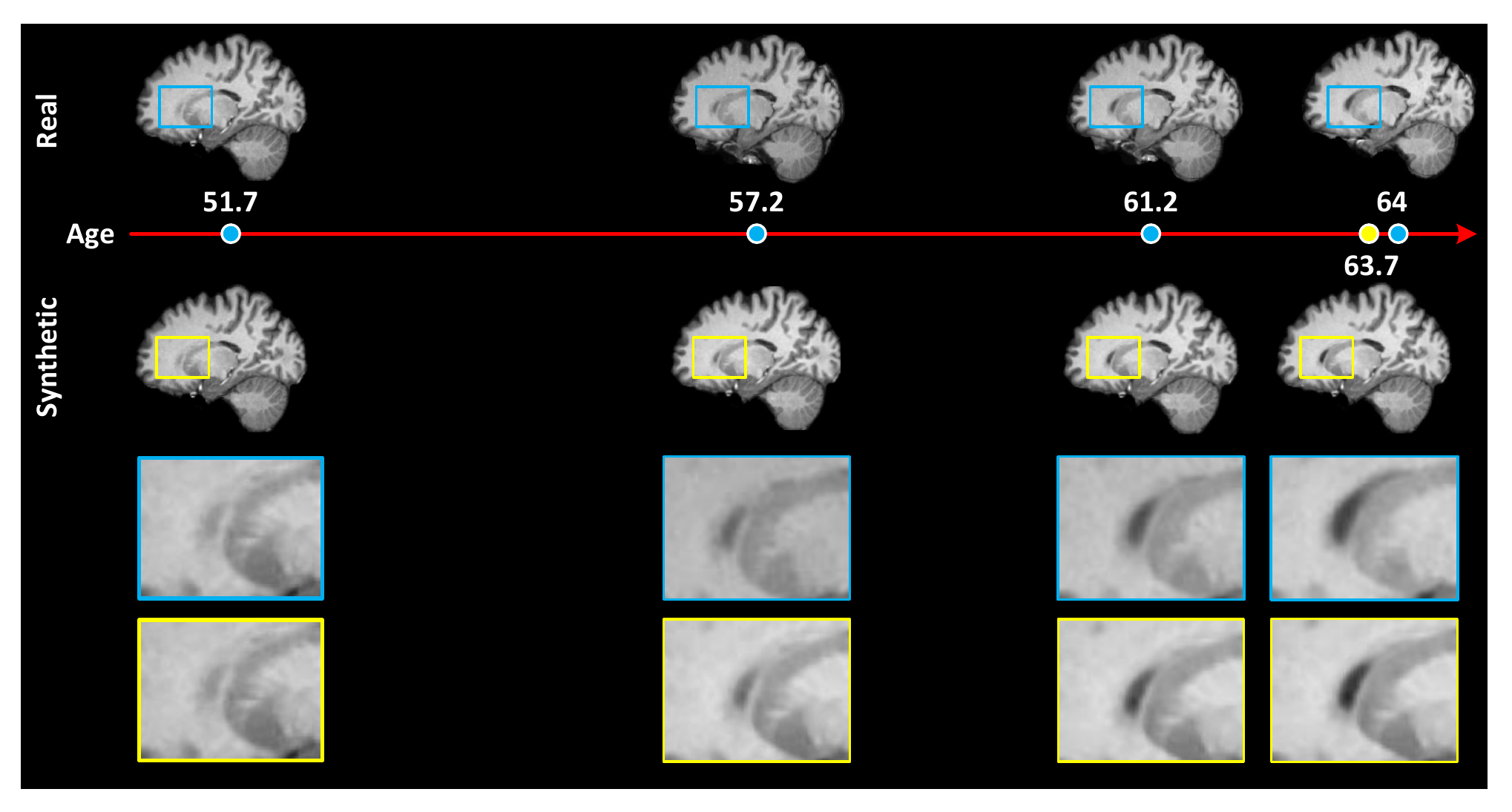}
\caption{Qualitative assessment of the quality of the synthetic MRI scans vs. the real MRI scans. The first row indicates the real MRI scans in the longitudinal data set, and the second row shows synthetic aging MRI scans at different estimated ages. We also show a magnified region at the bottom of the figure for each row respectively (color figure online). As the synthetic MRI scans are acquired along with the same interval of 6 months, the last obtained brain age is 63.7 in this case, whereas the counterpart in real data is 64. 
}
\label{fig:comparsion_real_syn}
\end{figure}

\begin{figure}
\centering
\includegraphics[width=1\linewidth]{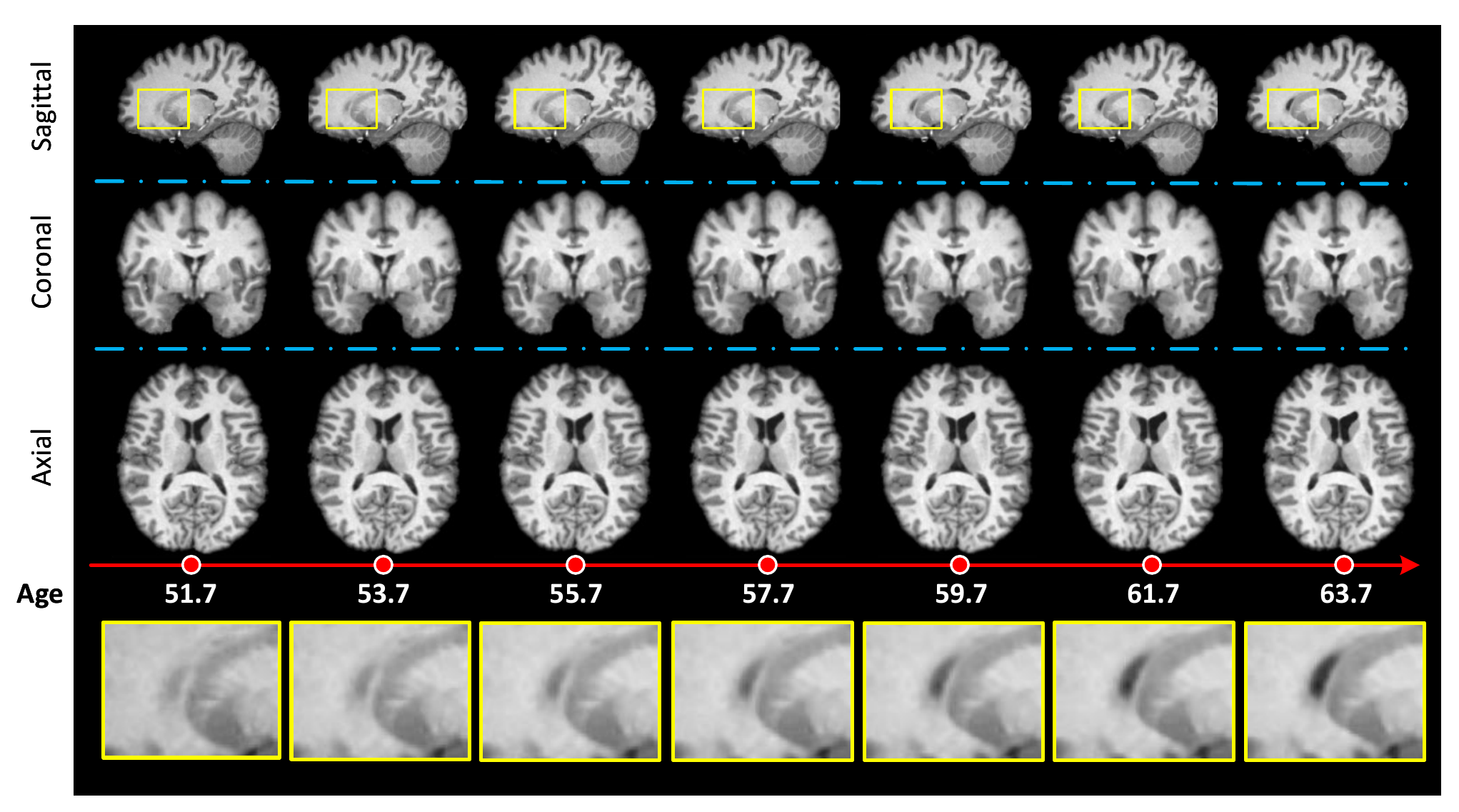}
\caption{The aging simulations were synthesized using our methodology for a healthy subject from age 51.7 to age 63.7. The 3D MRI scans are shown in three directions sagittal, coronal, and axial, respectively. A magnified region is highlighted at the bottom of the figure for a better illustration (color figure online).
}
\label{fig:AGING}
\end{figure}

%In this section, we evaluate the ability of aging simulation of the proposed method in two ways: 1) We show the simulated aging MRI scans from three datasets to show it from visualization; 2) we calculate the "middle" MRI scans' extreme age as \emph{linear age}, and introduce the Root Mean Square Error (RMSE) upon the \emph{linear ages} and \emph{true ages}.

%\textbf{Visualization of aging.} 
Figure.~\ref{fig:comparsion_real_syn} shows an example subject of the generated aging images. The comparison between real aging MRI scans and synthetic aging MRI scans shows the good quality of the synthetic images. %Firstly, we randomly select two samples from two datasets respectively, as shown in Figure.~\ref{fig:AGING}.
% As shown in Fig.~\ref{fig:AGING}, we randomly selected three samples each from three datasets.
Figure.~\ref{fig:AGING} shows the estimated aging progression of this subject, where the aging MRI scans are shown in three directions (coronal, sagittal, and axial), ranging from 51.7 years old to 63.7 years old. We report a magnified region to show the aging progression in the sagittal direction. %Each sample has a different age interval, i.e. 2 years for ADNI sample, 1 year for OASIS-3 sample. As shown in Fig.~\ref{fig:AGING}, we show the difference maps calculated between the real baseline and the corresponding aging MRI scans in each column. Red and blue show positive and negative, respectively. 
As shown, the ventricles expand with increasing age. This can be seen as an indication that the proposed method is consistent with what is expected in the aging brain. 

%Fig. \ref{fig:AGING} shows some slices of the real and synthetic images for a random subject per dataset. 
Although the synthetic images look realistic to untrained eyes, it is necessary to perform validation with a neuroradiologist in order to assess the quality of the generated images. This step is important towards the use of the synthetic data for answering clinical questions.

%Finally, we set up a task to assess the quality of the generated MRI scans from a clinical standpoint. 

In order to accomplish this, we designed a discrimination task for the neuroradiologist (A.T.), who has 16 years of experience with neurological images. The discrimination task consisted of distinguishing real images from synthetic ones. From the pool of mixed generated and real images, 200 MRI scans were randomly selected. Because we are interested in knowing if there was any bias in different datasets and age difference between the oldest and youngest image of the subject used to generate the image, we selected the 200 samples proportionally for each subcategory. By dividing the age difference into six ranges, (i.e., 2 years per range), we got [0$\sim$2,2$\sim$4,4$\sim$6,6$\sim$8,8$\sim$10,10+]. We included the generated MRI scans at time point $s$ as the generated one in the pool. 

\begin{minipage}{\textwidth}
        \begin{minipage}[h]{0.45\textwidth}
            \centering
            \makeatletter\def\@captype{table}\makeatother\caption{Confusion matrix on the discrimination task performed by the neuroradiologist.}
            \begin{tabular}{c|cc|ll}
            \cline{2-3}
                                            & \multicolumn{2}{c|}{\textbf{Neuroradiologist Assessment}}                                                                                        &  &  \\ \cline{2-3}
                                            & \multicolumn{1}{c|}{Real (P)}                                                     & Synthetic (N)                                                &  &  \\ \cline{1-3}
            \multicolumn{1}{|c|}{Real}      & \multicolumn{1}{c|}{\begin{tabular}[c]{@{}c@{}}True Positive\\ =41\end{tabular}}  & \begin{tabular}[c]{@{}c@{}}False Negative\\ =24\end{tabular} &  &  \\ \cline{1-3}
            \multicolumn{1}{|c|}{Synthetic} & \multicolumn{1}{c|}{\begin{tabular}[c]{@{}c@{}}False Positive\\ =49\end{tabular}} & \begin{tabular}[c]{@{}c@{}}True Negative\\ =86\end{tabular}  &  &  \\ \cline{1-3}
            \end{tabular}
            \label{table:confusion_m}
        \end{minipage}
        \begin{minipage}[h]{0.55\textwidth}
            \centering
            \includegraphics[height=0.7\textwidth]{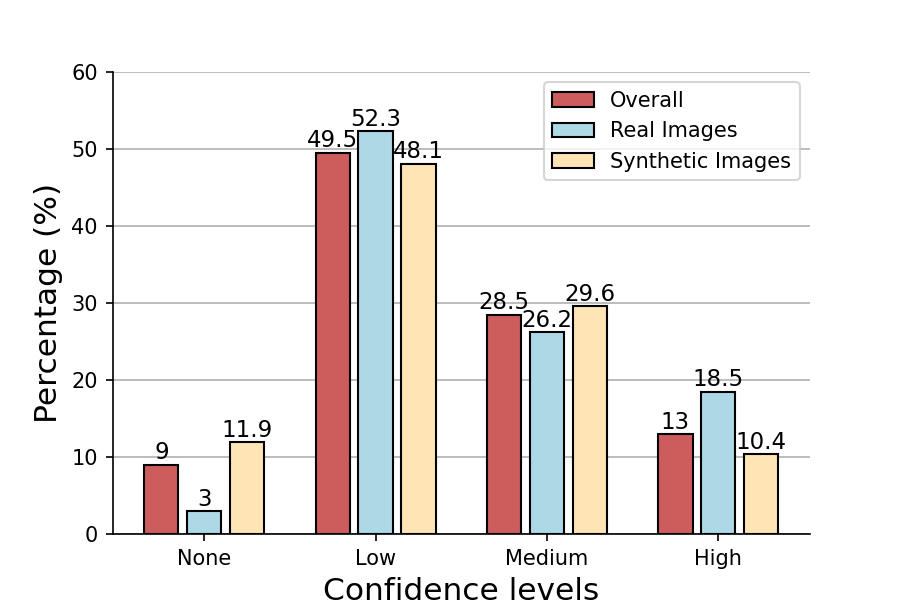}
            \makeatletter\def\@captype{figure}\makeatother\caption{The distribution of confidence levels}
            \label{figure:confidence_l}   
        \end{minipage}
\end{minipage}

\begin{table}[h]
\centering
\caption{Comparison of the accuracy and F1-scores of the assessment performed by the neuroradiologist.}
\label{table:metrics}
\resizebox{\textwidth}{13mm}
{% 
\begin{tabular}{c|c|cccc|ccc|cccccc}
\cline{1-15}
\multirow{2}{*}{\textbf{Criteria}} & \multirow{2}{*}{\textbf{Overall}} & \multicolumn{4}{c|}{\textbf{Confidence Levels}} & \multicolumn{3}{c|}{\textbf{Datasets}} & \multicolumn{6}{c}{\textbf{Age differences}  }                       \\ \cline{3-15} 
 &                          & None    & Low    & Medium    & High    & OASIS-3    & ADNI    & GENIC   & 0$\sim$2 & 2$\sim$4 & 4$\sim$6 & 6$\sim$8 & 8$\sim$10 & 10+ \\ \hline
\textbf{Accuracy}                            & 0.64 & 0.56 & 0.66 & 0.58 & 0.73 & 0.67 & 0.62 & 0.57 & 0.46 & 0.67 & 0.66 & 0.74 & 0.50 & 0.71 \\
\textbf{F1-score (overall)}   & 0.62 & 0.45 & 0.62 & 0.57 & 0.73 & 0.65 & 0.62 & 0.42 & 0.41 & 0.66 & 0.61 & 0.73 & 0.49 & 0.71 \\
\hline
\textbf{F1-score (real)} & 0.53 & 0.20 & 0.51 & 0.50 & 0.74 & 0.58 & 0.56 & 0.13 & 0.24 & 0.60 & 0.47 & 0.70 & 0.44 & 0.67 \\
\textbf{F1-score (synthetic)} & 0.70 & 0.69 & 0.73 & 0.64 & 0.72 & 0.72 & 0.67 & 0.71 & 0.58 & 0.72 & 0.75 & 0.77 & 0.55 & 0.75 \\
 \hline
\end{tabular}
}%
\end{table}

The neuroradiologist was completely blinded to the purpose and design of the study, as well as to the demographic and clinical characteristics of the study participants. The neuroradiologist used the 3D slicer \footnote{https://www.slicer.org/}  for classifying the 200 selected MRI scans as real or synthetic. He completed the task in four consecutive days -- one hour of assessment per day. Before doing the task, the neuroradiologist was exposed to three true MRI scans, one per dataset, in order to help the neuroradiologist to build a template of a real image in this dataset. %When an expert is performing the task, it is also important to gather difficult information. 
Motivated by the experiments by \citet{ravi2022degenerative}, for each case, the expert was asked to assign a confidence level from the given list:
\begin{itemize}
  \item None: 'I have no idea, I am guessing the class of this scan'
  \item Low: 'I have low confidence in my answer'
  \item Medium: 'I am reasonably confident in my answer'
  \item High: 'I am absolutely sure in my answer'
\end{itemize}

Table \ref{table:confusion_m}, shows the confusion matrix of the assessment in which the real class is marked as positive (P) and synthetic as negative (N).  As shown, highly skilled experts struggle to recognize the real and synthetic images with an \emph{accuracy} of 63.5$\%$. Specifically, while even the expert can reach an accuracy rate of 63.5$\%$, the precision for detecting real images (45.6$\%$) is poor, suggesting our synthetic images look realistic.

Figure \ref{figure:confidence_l} shows the distribution of confidence levels reported by the neuroradiologist in the experiment. As we can see, the expert has high confidence in only 13\% of the cases. %only Low, Medium, or even None confidence most of the time (i.e. 77$\%$ of the time) in regards to the answers for the discrimination task. 
Moreover,  the neuroradiologist has more uncertainty for synthetic images (i.e., None confidence in synthetic was 11.9$\%$ compared to 3$\%$ for real images, p-value = 0.042). %Those are yet another indication that it is difficult to distinguish between the synthetic and the real ones. 

Table \ref{table:metrics} shows the accuracy and F1-score of the neuroradiologist in distinguishing between real and synthetic images. 
Moreover, we independently report the F1-scores for the synthetic and real images in Table \ref{table:metrics}. 
%Additionally, since we care about the ability to distinguish both real and synthetic images, we introduce the \emph{weighted F1-score} (represented as the F1-score (overall) in Table~\ref{table:metrics}) which is calculated by averaging the F1-score for each side (real and synthetic). 
To evaluate if the algorithm shows differences among different datasets and confidence levels, we also calculate the metrics for each subcategory. It is also worthwhile to analyze the age difference between the youngest and oldest images from the subjects since in our experiments we chose to generate $N$ images for each subject based on such age difference.
According to Figure~\ref{fig:datasets}, there is a large age difference between the three datasets, so we decided to divide the range of age differences into six small intervals, each of which contains 2 years. 

%The results are shown in Table~\ref{table:metrics}. 
% {\color{red} Antonios could say something here?}  
As expected, the neuroradiologist was able to distinguish better the two classes when his confidence was high. Regarding confidence levels between none to medium, the performance was better for synthetic images for the same level of confidence, because the neuroradiologist tends to be more synthetic-oriented as demonstrated by the precision for detecting real images of only 45.6\%.
%Interestingly, even with None confidence, the expert can get high F1-score by 0.69 for synthetic images, which indicates that it is synthetic-oriented in this trial. 
The accuracy was similar for different datasets (62$\pm$5$\%$), whereas the F1-score on overall for GENIC is about 0.2 lower than the other two datasets. One possible reason for this is that images from GENIC come from younger subjects that might have fewer visual distinctions due to age. This can make the deformation fields to be small, which increases the chances of synthetic being more similar to the real ones. %have a contrast that is quite different from the other two datasets. Thus, this might make real images look as synthetic ones.

We conducted proportion hypothesis tests on the age differences in accuracy and corrected the resulting p-values by multiple comparisons (p < 0.05). Based on the results, we observed that there is no significant difference among subgroups even over a period of ten years, thus indicating the validity of QCM and robustness of the proposed method. 
% the numbers are relatively similar for age differences larger than two years. It appears that images with age differences below two years are more difficult to distinguish. This was expected, since the deformation fields used to create these synthetic images will be smaller.

%our method performs well across all age difference ranges (should exclude two outliers at 0$\sim$2 and 8$\sim$10 due to the limited support samples).

%In summary, from the assessment by the radiologist , our method i) can generate realistic MRI scans which are hard to be distinguished even by the expert; ii) can accurately provide predictions regardless of different datasets and different age differences.

% \begin{table}[h]
% \centering
% \caption{}
% \begin{tabular}{cccc}
% \cline{3-4}
%  &                                                                              & \multicolumn{2}{c}{\textbf{Predicted by radiologist}} \\ \cline{3-4} 
%  & \multicolumn{1}{c|}{\begin{tabular}[c]{@{}c@{}}Total Population\\ = 200\end{tabular}} & Real (P)                & Synthetic (N)               \\ \cline{2-4} 
% \multicolumn{1}{|c|}{\multirow{2}{*}{\rotatebox{90}{\parbox[c]{1cm}{\centering\textbf{Ground Truth}}}}} &
%   \multicolumn{1}{c|}{Real (P)} &
%   \begin{tabular}[c]{@{}c@{}}True Positive\\ = 41\end{tabular} &
%   \begin{tabular}[c]{@{}c@{}}False Negative\\ = 24\end{tabular} \\
% \multicolumn{1}{|c|}{} &
%   \multicolumn{1}{c|}{Synthetic (N)} &
%   \begin{tabular}[c]{@{}c@{}}False Positive\\ = 49\end{tabular} &
%   \begin{tabular}[c]{@{}c@{}}True Negative\\ = 86\end{tabular} \\ \cline{2-4} 
% \end{tabular}
% \end{table}

\section{Discussion}\label{section:Discussion}

We presented a method to efficiently generate 3D MRI scans of aging brains with the aim of augmenting the current longitudinal datasets with high-quality images. %that can provide insights into courses of the brain as it changes with age or disease. 
Our method is able to leverage DL-based methods to generate synthetic images while avoiding the time-consuming training process. In addition, we propose a series of strategies to ensure that the model can provide predictions of age for the generated images. 

\textbf{Image generation.}
The proposed method was evaluated on 2,662 T1-w MRI scans from 796 participants collected in three different datasets. While T1-w MRI is the only modality used, the methodology is applicable to any other modality.
We synthetized 7,548 high-quality images from the three datasets, which corresponds to an increase of 284\% in the number of images. Notice that the synthetic images were generated in the range between the youngest and oldest image per subject. It is actually straightforward to generate more images by extrapolating the images beyond the oldest one. In fact, we discarded approximately one-third of the synthetic images because we first generated images up to the integration point of three to estimate the stopping point $s$, which was close to two, so images beyond $s$ were discarded. The reason for this decision is that we were focused on creating a dataset of high-quality images that can be combined with existing datasets for brain aging analysis. While we think images beyond the stopping point $s$ are also of high quality, it is more difficult to assess that at this point. Future works may expand our current method in that direction.

Notice that we decided to exclude AD patients or subjects with mild cognitive impairment in this study. Indeed, it is straightforward to apply the method to these subjects. However, the aging patterns of these patient groups coexist with disease-related patterns, making the age estimation more difficult than in cognitively healthy individuals. This means that a synthetic dataset of aging in patients must consider this problem in order to make it useful for aging studies. Thus, we decided to develop and demonstrate the goodness of our method in healthy individuals, and future developments shall consider applying the method to disease populations such as Alzheimer´s disease.

%that method can be used to generate much more images by  we accept the images in the interpolation part and reject the extrapolation part, since the quality degrades in the extrapolation part, whereas we aim at maintaining quality. 

\textbf{QCM Verification.} Generally speaking, DL models always face a problem that the performance of the model is limited to the distribution of training data. When the distribution of the prediction data and the distribution of the training are different, the performance of the model tends to drop. We tackled that problem by using quality measurements to estimate the most similar generated image to the acquired ones. As shown in Fig.~\ref{fig:plot_s}, the closest image was almost always at a stopping point higher than one regardless of the used quality measurement, while theoretically that should be $1.0$. This way, we managed to benefit from the speedup provided by DL-based registration methods while keeping image quality as high as possible.

We selected six different quality measurements for estimating the stopping point $s$ and to estimate the quality of the images. %The recent GAN-based method \citep{heusel2017gans} also introduced a new evaluation criterion - Fréchet Inception Distance (FID) - to measure the similarity of synthetic and true images. In spite of its consistency with increasing disturbances and human judgment, FID suffers from a high bias in the number of samples used in the comparison~\citep{Binkowski2018DemystifyingMG}, and cannot detect overfitting~\citep{NEURIPS2018_e46de7e1}. More importantly, FID measures the difference between two distributions, whereas we need the measurement that can provide image-level measure and serve as a direct guide to updating the stopping point in the model.
All criteria gave consistent results as the closest synthetic image was always growing with the age of the acquired
image for all criteria (e.g., Fig. \ref{fig:curves}).

The results also show that these measurements perform similarly for age estimation, with NFN, PSNR, and NCC being slightly better (cf. Table \ref{tab:quantitative_measures}). Except for DSC, these measurements also show that the generated images are of good quality (cf. Table \ref{tab:quantitative_measures1}). However, the obtained DSC values are similar to the ones from previous studies, especially \citet{hoffmann2020synthmorph}. Moreover, the small size of some regions of the brain can be biased in the estimation of DSC.

While the focus of this paper has been to generate T1-w images, the method can be used for other modalities (e.g., T2w, FLAIR, etc.) without any change.
More interesting would be to generate images of one modality using images from another modality. For example, let's assume that in the first session, T1-w and T2w images were acquired, but only a T1-w was acquired in the second session. Then, the two T1-w images can be used to generate the deformation fields that can be applied to the only available T2w image. Another example would be if only a T1-w is available from the first session and only a T2w image is available for the second. In this case, it is necessary to use quality measurements that can deal with multimodality data, e.g., DSC or mutual information.

%and summarize four commonly used criteria: Mean Absolute Error (MAE), Structural Similarity Index (SSIM), NCC, and Peak Signal-to-Noise Ratio (PSNR). As another example, the Frobenius Norm (FN) is used since one of its main properties is that it is invariant under rotations and orthogonal transformations. These measurements can be used to measure similarity from an image-perspective. Yet, it is also important to measure similarity from a segmentation perspective. Therefore, we used the Dice Score Coefficient (DSC) whenever segmentation masks were available. 

%In particular, in cross-modality registration tasks, MI should be used since it is based on statistical information rather than spatial information, so it is insensitive to intensity; however, the NCC is preferable when mono-modality images are required, which is also the main emphasis of this paper.

\textbf{Qualitative assessment.} From the experiments, the synthetic images are of high quality. The assessment performed by the neuroradiologist (A.T.) shows that it is difficult to distinguish between synthetic images and real ones, especially when the neuroradiologist's confidence was not high. As expected, the images generated between smaller age differences were more difficult for the neuroradiologist, since the deformation fields in those cases are very small as well as age-related brain changes like small silent infarcts, enlarged perivascular spaces, cortical atrophy, white matter changes, and microbleed were absent. An interesting result was that the neuroradiologist tends to think that real images from GENIC look synthetic. Something that we have to consider is that the real images were preprocessed with Freesurfer (e.g., they are bias-corrected and skull stripped), which can make the visual assessment of the neuroradiologist slightly different compared with everyday clinical praxis in a neuroradiology department.

It is worth mentioning that in ~\citet{ravi2022degenerative}, the same discrimination task was introduced. They found that neuroradiologists can achieve an accuracy of $68.0\%\pm7.1\%$ on the synthetic images generated by their method, while our method can achieve a $64\%$ accuracy on the generated images. Although these numbers cannot be directly compared, this indicates that our performances might at least be comparable to the results in~\citet{ravi2022degenerative}. 

\textbf{Age estimation.} 
Based on the literature, we assumed that the brain changes due to age were linear. Such an assumption was supported by the regression analysis of Fig. \ref{fig:linear_fit}. After correcting the age estimation with the fitted lines, the age estimation was just slightly better (cf. Table \ref{tab:quantitative_measures} and Fig. \ref{fig:RMSE_box}). Notice that the research community has been very active in estimating age from images \citep{sajedi2019age, LUND2022102921}. Therefore, we foresee that the error in the age estimation of the generated images will decrease when using one of such methods. However, assessing different age estimation methods is beyond the aims of this study.

%In the section on age estimation, we accept the assumption that the brain changes linearly as it ages. Based on this assumption, we can assign linear ages to the augmented data. Even so, there are some potential ways to get a more accurate estimation of age. A simple solution would be to use a well-trained model to predict the target age and get a more accurate result. However, due to the high dimensionality of the 3D MRI scan, training such a model is quite expensive. We therefore hold this assumption and further verify the linear assumption is acceptable in our experiments.

\textbf{Future work.} We can see many potential applications and improvement directions for the proposed method. First, as mentioned, the method can be used to synthesize high-quality and high-resolution images from various modalities rapidly by only changing to the suitable measurement. Second, our method could be applied to estimate the progression of neurodegenerative brain diseases, not just for normal aging. For example, with our methodology, we can synthesize subject-specific temporal estimations of undergoing neurodegeneration, which can then be compared with the healthy templates to provide cross-sectional comparisons that shall aid clinical diagnoses.
%patient data can be compared to templates of normal subjects per age to estimate the so-called \textit{biological age} of the patient.
% One application we found is that we could observe the aging rate of patients with depression and compare it with those who have antidepressant treatment \cite{BALLESTER2021102864}. That can also be used to simulate different time courses for the patient's brain.
Finally, although our method is able to provide personalized predictions via paired inputs, it cannot be currently used when a single image is available from the subject. Our plan is to extend our framework to deal with that case as well. 

\section{Conclusion}\label{section:Conclusion}
In this work, we proposed a methodology with the aim of simulating subject-specific aging in brain MRIs given two 3D images acquired at different time points. DL-based diffeomorphic registration was used as a backbone to generate deformation fields at different integration points. Quality measurements were used for controlling the age estimation of the generated images by using a linear assumption. %The integration range was controlled and fine-tuned according to the quality of generated MRI images. 
The results show good performance from both quantitative and qualitative perspectives regarding both the image quality of the synthetic MRI scans and the estimation of age.

To help data-hungry AI models learn, we open-sourced  our augmented data to the research community, which is available in this link \url{https://github.com/Fjr9516/Synthetic-Brain-Aging/blob/main/README.md}.

\section*{Funding}  
This study has been partially funded by Vinnova through AIDA, project ID: 2108, by the China Scholarship Council (CSC) for 4 years’ study at KTH Royal Institute of Technology, by Digital Futures, project dBrain, by the Center for Innovative Medicine (CIMED), the regional agreement on medical training and clinical research (ALF) between Stockholm County Council and Karolinska Institutet, Hjärnfonden, Alzheimerfonden, Demensfonden, Neurofonden, Stiftelsen För Gamla Tjänarinnor, Fundación Canaria Dr. Manuel Morales (calls in 2012, 2014 and 2017); Fundación Cajacanarias; and the Estrategia de Especialización Inteligente de Canarias RIS3 from Consejería de Economía, Industria, Comercio y Conocimiento del Gobierno de Canarias, co-funded by the Programa Operativo FEDER Canarias 2014–2020 (ProID2020010063). The funders of the study had no role in the study design nor the collection, analysis, and interpretation of data, writing of the report, or decision to submit the manuscript for publication.
 
\section*{Acknowledgements}  

Data were provided in part by ADNI. Data collection and sharing for this project were funded by the Alzheimer's Disease Neuroimaging Initiative (ADNI) (National Institutes of Health Grant U01 AG024904) and DOD ADNI (Department of Defense award number W81XWH-12-2-0012). ADNI is funded by the National Institute on Aging, the National Institute of Biomedical Imaging and Bioengineering, and through generous contributions from the following: AbbVie, Alzheimer’s Association; Alzheimer’s Drug Discovery Foundation; Araclon Biotech; BioClinica, Inc.; Biogen; Bristol-Myers Squibb Company; CereSpir, Inc.; Cogstate; Eisai Inc.; Elan Pharmaceuticals, Inc.; Eli Lilly and Company; EuroImmun; F. Hoffmann-La Roche Ltd and its affiliated company Genentech, Inc.; Fujirebio; GE Healthcare; IXICO Ltd.; Janssen Alzheimer Immunotherapy Research \& Development, LLC.; Johnson \& Johnson Pharmaceutical Research \& Development LLC.; Lumosity; Lundbeck; Merck \& Co., Inc.; Meso Scale Diagnostics, LLC.; NeuroRx Research; Neurotrack Technologies; Novartis Pharmaceuticals Corporation; Pfizer Inc.; Piramal Imaging; Servier; Takeda Pharmaceutical Company; and Transition Therapeutics. The Canadian Institutes of Health Research is providing funds to support ADNI clinical sites in Canada. Private sector contributions are facilitated by the Foundation for the National Institutes of Health (www.fnih.org). The grantee organization is the Northern California Institute for Research and Education, and the study is coordinated by the Alzheimer’s Therapeutic Research Institute at the University of Southern California. ADNI data are disseminated by the Laboratory for Neuro Imaging at the University of Southern California. 

Data were provided in part by OASIS-3: Principal Investigators: T. Benzinger, D. Marcus, J. Morris; NIH P50 AG00561, P30 NS09857781, P01 AG026276, P01 AG003991, R01 AG043434, UL1 TR000448, R01 EB009352. AV-45 doses were provided by Avid Radiopharmaceuticals, a wholly owned subsidiary of Eli Lilly. 

The authors would like to thank Dr. Antonio Rodríguez for providing access to participants in GENIC and the Servicio de Resonancia Magnética para Investigaciones Biomédicas del SEGAI (University of La Laguna, Spain). Data used in preparation of this article is part of the GENIC-database (Group of Neuropsychological Studies of the Canary Islands, University of La Laguna, Spain. Principal investigator: Professor José Barroso. Contact: Dr. Daniel Ferreira, daniel.ferreira.padilla@ki.se). The following collaborators contributed to the GENIC-database but did not participate in analysis or writing of this report (in alphabetic order by family name): Nira Cedrés, Rut Correia, Patricia Diaz, Aida Figueroa, Nerea Figueroa, Eloy García, Lissett González, Teodoro González, Zaira González, Cathaysa Hernández, Edith Hernández, Nira Jiménez, Judith López, Cándida Lozano, Alejandra Machado, María Antonieta Nieto, María Sabucedo, Elena Sirumal, Marta Suárez, Manuel Urbano, and Pedro Velasco.

\section*{Author contributions}
\textbf{JF}: conceptualization; formal analysis; investigation; methodology; software; validation; visualization; writing - original draft; writing - review \& editing. 
\textbf{AT}: validation; investigation; writing - review \& editing.
\textbf{JB}: resources; data curation; writing - review \& editing.
\textbf{EW}: resources; data curation;  writing - review \& editing.
\textbf{DF}: conceptualization; methodology; resources; data curation; writing - review \& editing; supervision; funding acquisition.
\textbf{RM}: conceptualization; methodology; visualization; resources; project administration; writing - review \& editing; supervision; funding acquisition.

\bibliographystyle{unsrtnat}
\bibliography{main}  %%% Uncomment this line and comment out the ``thebibliography'' section below to use the external .bib file (using bibtex) .





% \end{thebibliography}

\end{document}